\documentclass[]{aa}

\usepackage{graphicx} 
\usepackage{rotating,subfigure,amssymb,afterpage} 
\usepackage{txfonts}
\usepackage[latin1]{inputenc} 
\usepackage{natbib}
\usepackage{supertabular}
\bibpunct{(}{)}{;}{a}{}{,}
\usepackage[pdftitle={}, 
pdfauthor={A. Reichert et al.}, 
pdfsubject={}, pdfkeywords={}, bookmarks, bookmarksopen,  
colorlinks=true, linkcolor=blue, citecolor=black, anchorcolor=blue, 
urlcolor=blue]{hyperref}  

\DeclareMathSymbol{*}{\mathbin}{symbols}{"01}								
\def\sun{\hbox{$_\odot$}}                                   
 
\def\ie{\textit{i.e.}}                                      
\def\eg{\textit{e.g.}}                                      

\begin{document} 
 
   \title{Observational constraints on the redshift evolution of\\ 
   X-ray scaling relations of galaxy clusters out to $z\!\sim\! 1.5$}

   \author{A. Reichert\inst{1}, H. B\"ohringer\inst{1}, R. Fassbender\inst{1}, M. Mühlegger\inst{1} } 
 
   \offprints{H. B\"ohringer, hxb@mpe.mpg.de} 
 
   \institute{$^1$ Max-Planck-Institut f\"ur extraterrestrische Physik, 
                 D 85748 Garching, Germany, {\tt hxb@mpe.mpg.de}\\ 
             } 
 
   \date{Submitted 09/03/2011}

\abstract 
{A precise understanding of the relations between observable X-ray properties of galaxy clusters and cluster mass is a vital part of the application of X-ray galaxy cluster surveys to test cosmological models. An understanding of how these relations evolve with redshift is just emerging from a number of observational data sets.}
{The current literature provides a diverse and inhomogeneous picture of scaling relation evolution. We attempt to transform these results and the data on recently discovered distant clusters into an updated and consistent framework, and provide an overall view of scaling relation evolution from the combined data sets.} 
{We study in particular the most important scaling relations connecting X-ray luminosity, temperature, and cluster mass (M--T, L$_X$--T, and M--L$_X$) combining 14 published data sets supplemented with recently published data of distant clusters and new results from follow-up observations of the XMM-Newton Distant Cluster Project (XDCP) that adds new leverage to efficiently constrain the scaling relations at high redshift.}
{We find that the evolution of the mass-temperature relation is consistent with the self-similar evolution prediction, while the evolution of X-ray luminosity for a given temperature and mass for a given X-ray luminosity is slower than predicted by simple self-similar models. Our best fit results for the evolution factor $E(z)^\alpha$ are $\alpha\!=\!-1.04\pm0.07$ for the M--T relation, $\alpha\!=\!-0.23^{+0.12}_{-0.62}$ for the L-T relation, and $\alpha\!=\!-0.93^{+0.62}_{-0.12}$ for the M--L$_X$ relation. We also explore the influence of selection effects on scaling relations and find that selection biases are the most likely reason for apparent inconsistencies between different published data sets.}
{The new results provide the currently most robust calibration of high-redshift cluster mass estimates based on X-ray luminosity and temperature and help us to improve the prediction of the number of clusters to be found in future galaxy cluster X-ray surveys, such as eROSITA. The comparison of evolution results with hydrodynamical cosmological simulations suggests that early preheating of the intracluster medium (ICM) provides the most suitable scenario to explain the observed evolution.}

\authorrunning{Reichert et al.} 
\titlerunning{Evolution of X-ray scaling relations} 
   \maketitle 
 

\section{Introduction} 
\label{sec:Introduction}

Galaxy clusters have become important probes for the study of the evolution of the large-scale structure and for the test of cosmological models (e.g. \citet{2001Natur.409...39B}, \citet{2001ApJ...560L.111H}, \citet{2002ARA&A..40..539R}, \citet{2003A&A...402...53S}, \citet{2005astro.ph..7013H}, \citet{2005RvMP...77..207V}, \citet{2009ApJ...692.1033V}, \citet{2009arXiv0909.3099M}, \citet{2010A&A...514A..32B}) and provide an interesting laboratory to study galaxy evolution. Nowadays, the application of galaxy cluster surveys to cosmological studies is limited mainly by a lack of understanding of galaxy cluster properties and the precise scaling relations of observables with cluster mass, in particular at higher redshifts. 

To date, X-ray observations provide the most reliable and detailed characterization of galaxy clusters and X-ray parameters and are most widely used in cosmological galaxy cluster studies for several reasons: (i) X-ray luminosity is tightly correlated to the cluster mass 
(\citet{2002ApJ...567..716R}, \citet{2009AA...498..361P}), (ii) bright X-ray emission is only observed for evolved clusters with a deep gravitational potential well, and (iii) the X-ray emission is highly peaked, minimizing projection effects. While great progress has been made in the characterizing clusters in X-rays at low redshifts (\citet{2005AA...441..893A}, \citet{2006ApJ...640..691V,2009ApJ...692.1033V}, \citet{2008AA...482..451Z}, \citet{2009AA...498..361P},  \citet{2009arXiv0909.3099M}, \citet{2010A&A...514A..32B}, \citet{2009arXiv0902.4890A}), the understanding of the evolution of the scaling relations towards higher redshifts is less clear. Results are now emerging that provide insight into the redshift range beyond $z = 1$, but the different available cluster samples provide inconsistent scenarios for the redshift evolution of the scaling relations. Moreover, it is not trivial to compare the different results since they have been partly derived for different cosmologies, for different definitions of the scaling radius, and compared with different schemes of the self-similar scaling laws to quantify the evolutionary trend. Therefore, this work makes an effort to put the different results into a uniform framework and combine the available data to study the evolutionary trend over the widest redshift baseline available.

The data sets used from the literature comprise the work of \citet{2010arXiv1006.3068A}, \citet{2009AA...498..361P}, \citet{2009arXiv0909.3099M}, \citet{2008AA...482..451Z,2007AA...467..437Z}, \citet{2008ApJ...680.1022H}, \citet{2007MNRAS.382.1289P}, \citet{2007AA...472..739B}, \citet{2007arXiv0710.5782O},  \citet{2006ApJ...640..691V}, \citet{2006MNRAS.365..509M}, \citet{2005AA...441..893A}, \citet{2005ApJ...633..781K}, and \citet{2004AA...417...13E} and we include the data of an additional 15 clusters from recent publications and the XMM-Newton Distant Cluster Project (XDCP, \citet{2008arXiv0806.0861F}, \citet{2005Msngr.120...33B}). The latter set of clusters significantly improve the statistics in the redshift range $z = 0.8 - 1.4$. 

The derived evolution results are compared to the findings of hydrodynamical cosmological simulations assuming different heating and cooling scenarios and therefore allow an investigation of the ICM thermal history.     

Tighter constraints on the evolution of the X-ray luminosity of clusters for a given mass also allow us to make refined predictions about the number of high redshift clusters to be observed in future X-ray surveys. We illustrate this in the context of the eROSITA mission \citep{2010SPIE.7732E..23P}. 

The paper is structured as follows. In Sect.\,\ref{sec:DataAndDataAnalysis}, we briefly describe the cluster samples used and the way in which we have transformed these public results into a unified framework. We also outline the theoretical expectations of the scaling relations and an estimate of the selection bias inherent to our combined cluster sample. The local scaling relations and the results on their evolutionary trend with redshift are given in Sect.\,\ref{sec:Results}. In Sect.\,\ref{sec:ImplicationsForTheThermalHistoryOfTheIcm},  these results are compared to the predictions of numerical simulation studies and the implications for the ICM heating scenario are discussed. In Sect.\,\ref{sec:ImplicationsForTheErositaClusterSurvey}, we outline the impact of our results on the number of clusters to be detected with eROSITA and Sect.\,\ref{sec:SummaryAndConclusions}, we provide a summary and our conclusions.  

Throughout the article, we adopt a $\Lambda$CDM cosmology with $(\Omega_{\Lambda},\Omega_{M},H_0,w)\!=\!(0.7,0.3,70\ \mathrm{km\ s^{-1}\ Mpc^{-1}},-1)$.

\section{Data and data analysis}
\label{sec:DataAndDataAnalysis}

\subsection{The cluster sample}
\label{sec:TheClusterSample}

\nocite{2007AA...472..739B} \nocite{2007AA...467..437Z} \nocite{2005AA...441..893A} \nocite{2004AA...417...13E}

\begin{table*}    
\begin{center}
\caption[Sample Overview]{Overview of the publications used to compile the combined cluster sample.} \label{tab:sample_overview}
\begin{tabular}{cccccc}
\hline

{\bf Publication}   & {\bf Acronym} & {\bf Survey} & {\bf Instrument} & {\bf Cluster \#} & {\bf z Range} \\

\hline
\citet{2010arXiv1006.3068A} & Andersson10 & SPT & XMM/Chandra & 9 & 0.4-1.1 \\
\citet{2009AA...498..361P}  & Pratt09 & REXCESS &XMM & 26 & $ \leq 0.2$ \\
\citet{2009arXiv0909.3099M} & Mantz09 & BCS, REFLEX&Chandra/ROSAT & 42 &$\leq 0.3$ \\ 
\citet{2008AA...482..451Z}  & Zhang08 & LoCuSS &XMM & 37 & 0.14-0.3 \\
\citet{2008ApJ...680.1022H} & Hicks08 & RCS & Chandra & 8 & 0.6-0.9 \\
\citet{2007MNRAS.382.1289P} & Pacaud07 & XMM--LSS &XMM & {\bf 13} & $ \leq 1.05$ \\
\citet{2007AA...472..739B} & Branchesi07 & archival &Chandra/XMM& 4 & 0.25-0.46\\
\citet{2007arXiv0710.5782O} & OHara07 & archival &Chandra &  26 & 0.29-0.82\\ 
\citet{2007AA...467..437Z} & Zhang07 & pilot LoCuSS &XMM & 4 & 0.27-0.3 \\
\citet{2006ApJ...640..691V} & Vikhlinin06 & archival &Chandra/ROSAT & {\bf 2} & $ \leq 0.2$ \\
\citet{2006MNRAS.365..509M} & Maughan06 & WARPS & XMM/Chandra & 8 & 0.6-1 \\
\citet{2005AA...441..893A} & Arnaud05 & archival &XMM &  5 & $ \leq 0.15$ \\
\citet{2005ApJ...633..781K} & Kotov05 & archival &XMM & 5 & 0.4-0.7 \\
\citet{2004AA...417...13E} & Ettori04 & archival &Chandra & 28 & 0.4-1.3 \\

\hline
\end{tabular}
\end{center}
\end{table*}

The evolution of galaxy cluster X-ray scaling relations is investigated by means of a combined cluster sample compiled from a number of recent publications. To enable constraints on the redshift evolution of scaling relations, the clusters were selected to cover a wide redshift range from local systems out to $z\!=\!1.46$. We attempted to avoid the complications caused by the expected deviation of galaxy groups from the scaling laws for more massive clusters by applying an ICM temperature threshold of $T_\mathrm{min}\!=\!2$ keV.

Table\,\ref{tab:sample_overview} shows the publications used to compile the combined cluster sample. The acronyms listed there are used throughout the remainder of this paper when referring to the source publications. The derived X-ray properties of clusters included in more than one subsample are compared in Appendix\,\ref{sec:ComparisonOfClusterPropertiesForSystemsIncludedInMoreThanOneSubsample}. For these systems, the results derived with an analysis scheme that is most similar to ours, hence requiring the least corrections are used for the combined sample. Selection biases and the intrinsic scatter in the cluster population about the mean relations ensure that a determination of scaling relation evolution in the redshift range up to $z\!=\!0.8$ is challenging (see Sect.\ref{sec:SelectionBiasEstimate}). Therefore, an extensive sample of high redshift clusters is crucial for this study. For this purpose, in addition to the subsamples listed in Table \ref{tab:sample_overview}, high redshift clusters detected in the XDCP \citep{2008arXiv0806.0861F}, XMM--LSS \citep{2004JCAP...09..011P}, XCS \citep{2002ASPC..268...43R} surveys, the 2XMM catalogue \citep{2009AA...493..339W}, and by the South Pole Telescope (SPT) (Andersson10, \citet{2011arXiv1101.1286F}) were included in the combined sample. 

Table\,\ref{tab:clusterlist} lists the galaxy clusters included in the combined sample consisting of 232 systems, of which 40 are at $z\!>\!0.8$. The cluster temperature, X-ray luminosity, and mass used in the present analysis are listed in the table. The sample covers a range of cluster masses from about $5*10^{13} M_{\mathrm{\sun}}$ to $3*10^{15} M_{\mathrm{\sun}}$. For the distant cluster sample, cluster masses derived by means of the $\mathrm{Y}_X$--M relation were not considered owing to their dependence on the $\mathrm{Y}_X$--M scaling relation and its redshift evolution. The $z\!>\!0.3$-clusters used to place constraints on scaling-relation evolution cover an approximately uniform luminosity range over the entire redshift interval $z=0.3-1.46$ (see Appendix\,\ref{sec:LXZDistributionOfTheClusterSample}) and by construction, the distant cluster sample shows no strong morphological selection bias.

\subsection{Scaling theory}
\label{sec:ScalingTheory}

Assuming that the evolution of the ICM during cluster formation is governed solely by gravitational processes, clusters are expected to be self-similar objects whose X-ray properties and masses are connected by scaling relations predicted by the self-similar model (\eg \ \citet{1986MNRAS.222..323K}). Since galaxy clusters have no clearly defined natural outer boundary, a fiducial radius within which cluster properties are considered has to be chosen. Scaling theory is used to define this radius in such a way that it describes the same corresponding boundary for clusters of all sizes in the framework of the self-similar cluster structure model. In accordance with the homogeneous spherical collapse model of \citet{1972ApJ...176....1G} and detailed N-body simulations (\eg \ \citet{1995MNRAS.275...56N}), the fiducial radius is defined to enclose a spherical region with a mean overdensity of $\Delta$ times the critical density of the Universe $\rho_\mathrm{crit}(z)$ 

\begin{equation}
r_\Delta^3=\frac{3 M(r<r_\Delta)}{4\pi \rho_\mathrm{crit}(z) \Delta}.
\label{eq:rdelta}
\end{equation}

\noindent While this first-order self-similar model seems to describe the structure of dark matter haloes fairly well, additional gas physics including heating and cooling processes are needed to explain the ICM structure and the resulting X-ray properties. Consequently, the local scaling relations have been found to differ from the self-similar predictions in some cases (see \eg \ Pratt09), \eg \ the L$_X$--T  relation is steeper than expected. 

The self-similar model also predicts the evolution of scaling relations with redshift or lookback time. However, there are different schemes for defining the fiducial radius in order to enclose self-similar regions for clusters at different redshifts. One approach based on a spherical top-hat collapse model assumes that a galaxy cluster as it is observed has only recently formed at the given redshift and proposes adopting a redshift-dependent density contrast $\Delta_z$ when defining fiducial radii, where $\Delta_z$ can be expressed in terms of the density contrast at the virial radius at the cluster redshift

\begin{equation}
\Delta_z=\Delta(z=0)\frac{\Delta_{vir}(z)}{\Delta_{vir}(z=0)}.
\label{eq:deltazvir}
\end{equation}

\noindent \citet{1998ApJ...495...80B} give an expression for $\Delta_{vir}(z)$ in a flat $\Lambda$CDM-cosmology of

\begin{equation}
\Delta_{vir}(z)=18 \pi^2 +82[\Omega_m(z)-1]-39[\Omega_m(z)-1]^2,
\label{eq:deltavir}
\end{equation}

\noindent where $\Omega_m(z)\!=\!\Omega_M(1+z)^3/E(z)^2$ and $E(z)=\frac{H(z)}{H_0}$. The expectation for the M--T, L$_X$--T, and M--L$_X$ relation and their evolution with redshift in a model only taking into account gravitational effects is then

\begin{eqnarray}
M \propto T^{3/2}E(z)^{-1}\Delta_z^{-1/2}\label{mtevol},\\
L_X \propto T^2 E(z) \Delta_z^{1/2}\label{ltevol},\\
M \propto L_X^{3/4}E^{-7/4}\Delta_z^{-7/8},\label{mlevol}
\end{eqnarray} 

\noindent where $L_X$ is the bolometric luminosity integrated out to the scale radius. This or related definitions of fiducial radii were used in a number of recent publications on scaling relation evolution (\eg \ \citet{2004AA...417...13E} and \citet{2006MNRAS.365..509M}). The second commonly used definition is to measure cluster properties within regions with a redshift-independent value for the density contrast $\Delta$, which was applied \eg \ by \citet{2007MNRAS.382.1289P} and \citet{2005ApJ...633..781K}. The expected evolution of scaling relations in this framework is similar to Eqs.\,\ref{mtevol}, \ref{ltevol}, and \ref{mlevol}, albeit omitting the $\Delta_z$-factors. We are currently unable to decide between the two approaches on the basis of observational data owing to the lack of sufficiently extensive and precise data sets needed to catch the subtle differences between the two models. In a forthcoming paper (B\"ohringer et al. 2011), we explore this question by means of numerical simulations, finding that the recent formation approximation is imprecise and that the fixed overdensity approach (not including the $\Delta_z$ factors) describes the simulation results more accurately than the formulae including this extra term. For this paper we decided to explore both approaches and find that the differences are negligibly small for our conclusions.

\subsection{Homogenization scheme}
\label{sec:HomogenizationScheme}

A number of corrections had to be applied to the subsamples in order to correct for the slightly different methods used by the authors. 

Throughout the cluster samples, we used both means of defining fiducial radii described in the previous section. Within these two schemes, various values of the mean overdensity are used. Most recent studies use a density contrast of 200, 500, or 2\,500. The use of $r_{200}$, which approximately corresponds to the virial radius, has the drawback that in many cases the area inside $r_{200}$ is not fully covered by the available X-ray data. In many clusters, $r_{2500}$ typically corresponds to the most relaxed central part of the cluster and is also used in some publications. The most common definition of the cluster radius that is also used in this work is $r_{500}$ (corresponding to either $\Delta_z=500*\Delta_{vir}(z)/\Delta_{vir}(z=0)$ or a fixed $\Delta=500$). A large fraction of the clusters are well-relaxed within this radius, which is often well matched with the cluster region for which X-ray data are available.

A rescaling scheme similar to the one used by \citet{2007AA...472..739B} was used to correct for the different definitions and values of density contrast. For this rescaling scheme, the cluster density profiles were assumed to follow the isothermal $\beta$-model \citep{1976AA....49..137C}, since for the majority of the clusters included in the combined sample the shape parameters $\beta$ and $r_c$ are known. In the framework of the $\beta$-model, the mass enclosed within the radius $r$ is given by 

\begin{equation}
M(r)=\frac{3\beta k_B T }{G \mu m_p}\frac{r(r/r_c)^2}{1+(r/r_c)^2}.
\label{eq:Mbeta}
\end{equation} 

\noindent Using Eqs.\,\ref{eq:Mbeta} and \ref{eq:rdelta}, the radius corresponding to a density contrast of $\Delta_z$ is given by

\begin{equation}
r_{\Delta_z}=\sqrt{\frac{6\beta k_B T}{\mu m_p}\frac{1}{H(z)^2 \Delta_z}-r_c^2}. 
\label{eq:rbeta}
\end{equation}
 
 \noindent For the $\beta$-model, the fractional luminosity within $r$ is 

\begin{equation}
L_X(<r)/L_\mathrm{tot}=\frac{2\pi S_0}{3f_\mathrm{tot}} \left[ \frac{1-(1+(r/r_c)^2)^{3/2-3 \beta}}{2\beta -1} \right],
\label{eq:Lbeta}
\end{equation}

\noindent where $S_0$ designates the central surface brightness and $f_\mathrm{tot}$ the total X-ray flux. Using these equations, a simple correction scheme can be applied to the cluster observables. First $x_1\!=\!r_1/r_c$, the radius in units of $r_c$ for which the cluster properties are given is calculated using Equ.\,\ref{eq:rbeta}, where $\Delta_z$ is calculated by means of Equ.\,\ref{eq:deltazvir}. The radius corresponding to the density contrast to which the cluster properties will be rescaled,  $x_2\!=\!r_2/r_c$, is calculated in the same way. Using Eqs.\,\ref{eq:Mbeta} and \ref{eq:Lbeta} we obtain the expressions 

\begin{eqnarray}
M(x_1)/M(x_2)=\left( \frac{x_1^3}{1+x_1^2}\right) / \left( \frac{x_2^3}{1+x_2^2} \right) \label{eq:mfactor}\\
L_X(x_1)/L_X(x_2)=\frac{1-(1+x_1^2)^{3/2-3\beta}}{1-(1+x_2^2)^{3/2-3\beta}}. \label{eq:lfactor}
\end{eqnarray}
 
\noindent The cluster properties are then multiplied by the correction factors obtained by means of Eqs.\,\ref{eq:mfactor} and \,\ref{eq:lfactor}.

Some publications give cluster X-ray luminosities in either the 0.1-2.4 keV or 0.5-2 keV energy band. For the combined cluster sample, bolometric X-ray luminosities, that is cluster luminosities in the 0.01-100 keV-band, are used. The band luminosities were converted to bolometric values by means of the X-ray spectral fitting package XSPEC \citep{1985A&AS...62..197M}, assuming a Mekal model with an ICM metallicity of $0.3 Z_{\sun}$.

It has been shown (\eg \ \citet{1998ApJ...504...27M}, Pratt09) that tighter scaling relations involving X-ray luminosity are obtained by excising the cluster core region. For the combined sample selected in this work, however, the cluster observables for the local sample are given in a way that allows direct comparison to distant clusters for which core excision is not always feasible. Therefore, luminosities in the entire $r\!<\!r_{500}$ aperture are considered, the emission from central regions is not excluded or replaced by any extrapolated profile to make the results of local samples comparable to the distant cluster studies. In the case of the ICM temperatures, obtaining a homogeneous dataset is less straightforward because in some studies of local clusters only core-excluded temperatures are given. Compiling a combined sample directly from studies of clusters where the core is included and others with core-excluded temperatures can lead to systematic bias. The source of this bias is the diversity of the central ICM temperature profiles that affects the difference between core-included and core-excised temperatures. In most cases, non-CC clusters have a flat central temperature profile but CC clusters with a rather pronounced cool core (CC) have a central temperature is between about one-third and one-half of the surrounding regions (\eg \ \citet{2003ApJ...590..207P}). The fraction of CC clusters therefore determines the magnitude of the induced bias.

Subsamples with both core-included and core-excised temperatures available allow us to estimate the errors caused by the inhomogeneous measurement schemes. The sample of Pratt09 consists of 31 low redshift clusters and was designed to be morphologically unbiased, hence its relative error introduced by using core-excluded instead of core-included temperatures was found to be $7\%$. This value was added to the estimated temperature errors for samples with only core-excluded temperatures available (Andersson10, Zhang08, Zhang07, Arnaud05). For the cluster SPT-CL J2106-5844 \citep{2011arXiv1101.1286F}, an emission-weighted core-included temperature of $T\!=\!8.5$ keV was estimated from the core-excluded temperature of 11.0 keV and the core temperature of 6.5 keV. As outlined above, the importance of correct and homogeneous core exclusion throughout the combined sample depends on the abundance of CC clusters. For local systems, this abundance is found to be $40\!-\!70\%$ (\eg \ \citet{2010A&A...513A..37H}). No consensus has emerged yet on how this CC fraction evolves with redshift. While \citet{2009ApJ...692.1033V} find a decrease in the CC fraction from $\sim\!70\%$ locally to $\sim \!15\%$ at $z\!>\! 0.5$, \citet{2008ASPC..399..375S} find that the fraction of CC clusters at high redshift ($z\!\sim\!0.7-1.2$) is very similar to the local value (with an absence of very strong CCs). In summary, using inhomogeneous temperature measurement schemes throughout the combined sample is believed to bias the evolution results only marginally.
     
The values of most cluster properties depend on the assumed cosmological model. In all publications used to compile our combined cluster sample, the model of choice is the standard $\Lambda$CDM scenario. However, the  value assumed for the Hubble constant $H_0=100h\ \mathrm{km \ s^{-1} \ Mpc^{-1}}$ varies slightly from $h=0.7$ to $h=0.73$ within the list of source publications considered. The effects of this change slightly affect the values of the basic cluster properties. To obtain comparable cluster subsamples, a common $h$ of 0.7 is chosen and the necessary corrections are applied to the subsamples with a different $h$.

Cluster mass depends on $h$ as

\begin{equation}
M \propto h^{-1},
\end{equation}

\noindent while for the bolometric X-ray luminosities 

\begin{equation}
L_X\propto d_L^2\propto h^{-2}.
\end{equation}

\noindent The cluster properties were rescaled accordingly, \eg \  $M_{70}/M_{73}=73/70=1.043$ and {\bf $L_{70}/L_{73}=(73/70)^2=1.088$}.

\subsection{Selection bias estimate}
\label{sec:SelectionBiasEstimate}

The cluster samples obtained in typical X-ray surveys are not strictly volume-limited but rather, at least approximately, flux-limited since the limited observation time, and detector area and sensitivity generally only enables the detection of objects brighter than a certain flux limit $f_\mathrm{min}$. Various selection biases complicate the analysis of these flux-limited samples and have to be taken into account when determining scaling relations and their evolution with redshift. For a well-controlled, homogeneous survey, the cluster selection function, \ie \ the probability of detecting a cluster with given properties taking into account the adopted observation strategy, can be modeled. A realistic selection function enables us to correct for selection effects in a more consistent and exact way than a survey for which only an approximate flux limit is known. An example of the application of this correction strategy can be found in \citet{2007MNRAS.382.1289P}. However, the sample used in our work is highly heterogeneous, including clusters from numerous different surveys. For the majority of these surveys, not all information necessary to reconstruct the survey selection function with high accuracy is available. Therefore, a different strategy has to be applied to obtain at least a realistic estimate of the influence of selection effects for this sample.
 
The approach adopted here consists of simulating a cluster population with as realistic as possible properties and selecting a cluster sample comparable to the observed one from this population. In this situation, in contrast to observed cluster samples, the characteristics of the underlying cluster population are known and the properties of the selected sample can be compared to those of the entire population to estimate selection effects.
To probe the selection bias in the local scaling and the evolution of the L$_X$--T  relation, a temperature function $n_T(T,z)$ was assumed. Since no sufficiently exact measured temperature function was available, $n_T(T,z)$ was deduced from the more tightly constrained luminosity function $n_L(L,z)$ by multiplying it with the determinant $\mathrm{d}L/\mathrm{d}T$ and converting X-ray luminosities to temperatures by means of the L$_{0.1-2.4 \mathrm{keV}}$--T  relation. This method is only exact for the unrealistic case of no intrinsic scatter about the scaling relation, but provides a sufficiently exact approximation of the cluster temperature function to obtain a rough estimate of the selection bias. We note that for cluster discoveries and the related selection effects the luminosities in the X-ray observatory's detection band rather than the bolometric X-ray luminosities used throughout this work are relevant. Since the majority of combined sample clusters originate from the ROSAT surveys, ROSAT band luminosities (0.1 - 2.4 keV observer frame) are used throughout this section. The number of clusters in a given redshift bin is therefore determined by the temperature function and the solid angle covered by the survey.

Cluster luminosities were calculated assuming a non-evolving L$_{0.1-2.4 \mathrm{keV}}$--T  relation since this evolution model represents a fair first-order approximation to the observational data. The ROSAT band L$_{0.1-2.4 \mathrm{keV}}$--T  relation given in Pratt09, $L_{0.1-2.4 \mathrm{keV}}\!=\!0.078*T[\mathrm{keV}])^{2.24} 10^{44}\mathrm{erg\ s^{-1}}$, was used for this purpose. The luminosities were then displaced from the mean relation assuming a log-normal scatter of $0.25$ dex ($\sim\!60\%$), as suggested by the observed population and consistent with the value found by Pratt09. The chosen description is consistent with observations and seems reasonable when assuming a cluster's mass to be its basic property and taking into account the rather tight correlation between mass and temperature compared to the greater intrinsic scatter in the L$_X$--T  relation. As for the observed cluster sample, a temperature threshold of $T_\mathrm{min}\!=\!2$ keV was applied to the simulated population.

From this simulated cluster population, a flux-limited sample can be extracted including only clusters brighter than the limiting luminosity $L_\mathrm{min}(z)$ resulting from the flux limit as 

\begin{equation}
L_\mathrm{min}(z)=4\pi d_L^2 f_\mathrm{min}/K(z,T),
\label{flim}
\end{equation}

\noindent where $d_L$ is the luminosity distance and $K(z,T)$ the k-correction quantifying the relation between observer-frame band luminosity and cluster rest-frame band luminosity, \ie \ $K(z,T)=L_\mathrm{obs}/L_\mathrm{rest}$. 

The influence of selection effects on the observed local scaling relations was estimated by means of flux-limited samples selected from the simulated cluster population with limiting fluxes of $f_\mathrm{min}\!=\!3*10^{-12} \mathrm{\ erg\ s^{-1}\ cm^{-2}}$ (sample\,1), $1*10^{-12} \mathrm{\ erg\ s^{-1}\ cm^{-2}}$ (sample\,2), and $1*10^{-13}  \mathrm{\ erg\ s^{-1}\ cm^{-2}}$ (sample\,3), covering the redshift range used to fit the local relations, \ie \ $0\!<\!z\!<\!0.3$. These flux limits approximately represent the range of limiting fluxes of the local cluster surveys used in this work. The survey areas were adjusted to provide sample sizes comparable to both each other and the local cluster sample used in this work, \ie \ $\sim\!100$ clusters. The bolometric X-ray luminosities of the cluster population were assumed to follow the input L$_X$--T relation of Equ.\,\ref{ltloc}. We then fitted L$_X$--T  relations to the three samples using the BCES(L$|$T) method \citep{1996ApJ...470..706A}. Table\,\ref{ltsimtab} summarizes the derived relations. 

\begin{table}[t]    
\begin{center}
\caption{L$_X$--T  relations derived from the simulated local samples. Sample\,1 $f_\mathrm{min}\!=\!3*10^{-12} \mathrm{\ erg\ s^{-1}\ cm^{-2}}$, sample\,2 $f_\mathrm{min}\!=\!1*10^{-12} \mathrm{\ erg\ s^{-1}\ cm^{-2}}$, sample\,3 $f_\mathrm{min}\!=\!1*10^{-13} \mathrm{\ erg\ s^{-1}\ cm^{-2}}$. The input relation used to model the bolometric X-ray luminosities of the simulated cluster population is given for comparison.}\label{ltsimtab}
\begin{tabular}{cc}
\hline
input relation & $L_X\!=\!0.112* (T[\mathrm{keV}])^{2.53} 10^{44}\mathrm{erg\ s^{-1}}$ \\
\hline
{\bf sample}   & {\bf L$_X$--T  relation} \\
\hline
1 & $L_X\!=\!(0.20\pm 0.03)* (T[\mathrm{keV}])^{2.31\pm 0.12} 10^{44}\mathrm{erg\ s^{-1}}$\\
2 & $L_X\!=\!(0.22\pm 0.05)* (T[\mathrm{keV}])^{2.13\pm 0.19 } 10^{44}\mathrm{erg\ s^{-1}}$\\
3 & $L_X\!=\!(0.09\pm 0.02)* (T[\mathrm{keV}])^{2.78\pm 0.28 } 10^{44}\mathrm{erg\ s^{-1}}$\\
\hline
\end{tabular}
\end{center}
\end{table}

In summary, the characteristics of the selection effects for local scaling-relation fits depend on whether the flux limit of the sample cuts away a significant fraction of the luminosity function. For the faintest of the three samples described above (sample 3), this is not the case, and consequently selection bias is negligible for this sample. Most of the clusters within the $z\!<\!0.3$ sample used in this work were detected in surveys with relatively high flux limits, making the flux limit and not the applied temperature threshold the limiting factor at the low temperature and luminosity end of the cluster distribution. The bias in the observed sample is therefore expected to be comparable to the $f_\mathrm{min}\!=\!1*10^{-12}  \mathrm{\ erg\ s^{-1}\ cm^{-2}}$ and the $f_\mathrm{min}\!=\!3*10^{-12}  \mathrm{\ erg\ s^{-1}\ cm^{-2}}$-sample. As is clearly visible in the L$_X$--T  relations fitted to these samples, the resulting bias for this range of flux limits is fairly insensitive to the exact limiting flux. This situation justifies a common bias estimate for the entire local sample used in this work. According to the simulated samples, the measured slope of the local L$_X$--T  relation is decreased only slightly by selection bias, whereas the normalization is raised by almost $100\%$. The selection bias in the measured evolution of the L$_X$--T  relation with redshift displays similar trends, \ie \ a higher normalization caused by selection effects, and is analyzed in greater detail throughout the remainder of this section.

\begin{figure}[t]
	\flushleft
		\includegraphics[angle=-90,width=0.5\textwidth]{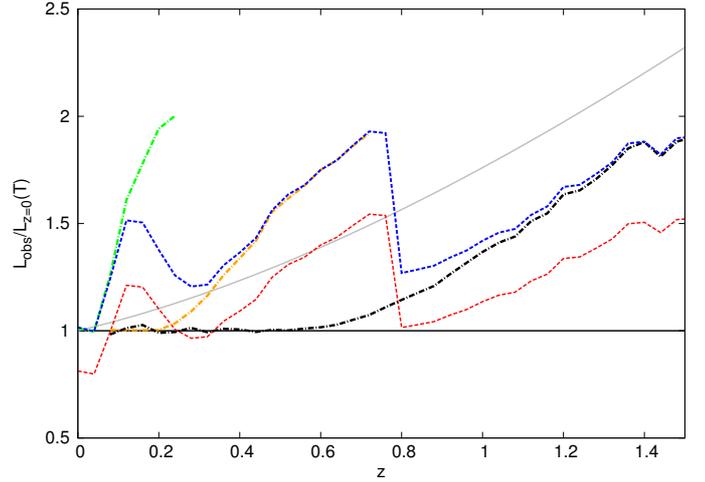}
	\caption{Evolution bias of the L$_X$--T  relation: Flux-limited samples with $f_\mathrm{min}\!=\!3*10^{-12} \mathrm{\ erg\ s^{-1}\ cm^{-2}}$ (green), $f_\mathrm{min}\!=\!1*10^{-13} \mathrm{\ erg\ s^{-1}\ cm^{-2}}$ (orange), $f_\mathrm{min}\!=\!1*10^{-14} \mathrm{\ erg\ s^{-1}\ cm^{-2}}$ (black), and combined sample (blue curve). The bias for the combined sample rescaled to remove the bias in the local relation is plotted in red. The self-similar prediction for the evolution is plotted in grey.}
	\label{fig:ltzbias}
\end{figure}

The simulated counterpart of the combined cluster sample used in this work was constructed from an all-sky survey with a flux limit of $f_\mathrm{min}\!=\!3*10^{-12} \mathrm{\ erg\ s^{-1}\ cm^{-2}}$, representing the cluster samples based on the ROSAT All-Sky Survey (RASS), \eg \ the REFLEX \citep{2004AA...425..367B} survey. A second sample with a flux limit of $f_\mathrm{min}\!=\!1*10^{-13} \mathrm{\ erg\ s^{-1}\ cm^{-2}}$ and an area of 400 square degrees was added, representing the clusters from ROSAT PSPC-based surveys such as WARPS \citep{2002yCat..21400265P} or the 400sd \citep{2007ApJS..172..561B} survey. As a third component, a sample with $f_\mathrm{min}\!=\!1*10^{-14} \mathrm{\ erg\ s^{-1}\ cm^{-2}}$, an area of 80 square degrees and a minimum redshift of $z_\mathrm{min}\!=\!0.8$ was added, corresponding to current serendipitous surveys such as XDCP \citep{2008arXiv0806.0861F}.

With this simulated sample at hand, the influence of bias on the evolution of the L$_X$--T  relation can be estimated. To achieve this, we calculated the logarithmic mean of the bolometric cluster luminosity divided by the luminosity resulting from the bolometric L$_X$--T  relation at the cluster temperature in redshift bins. This number quantifies the selection bias, with a value of 1 corresponding to no selection effects. The value of the bias curve in a redshift bin with a width of $\Delta z$ and centered on the redshift $z$ is therefore given by\\ 

\begin{equation}
B(z)=10^{\frac{\sum_{\mathrm{clusters}}\mathrm{log}\left(\frac{L_X}{A*T^b}\right)}{N_{\mathrm{clusters}}}},
\label{biascurve}
\end{equation}

\noindent where the subscript "clusters" designates all clusters within $[z-\frac12\Delta z;\ z+\frac12\Delta z]$ included in the flux-limited sample and $A$ and $b$ quantify the normalization and slope of the local L$_X$--T  relation and were set according to Equ.\,\ref{prattltloc}.

The bias curve deduced for the simulated combined sample is indicated with a blue-dashed line in Fig.\,\ref{fig:ltzbias}. However, in the redshift range used to fit "local" scaling relations, $0.05\!<\!z\!<\!0.3$, selection effects are already non-negligible, visible in a clear deviation of the bias curve from 1. A value of 1 on the vertical axis, theoretically corresponding to no bias, therefore already includes the bias present in the local sample used to fit the L$_X$--T  relation. This effect is redshift-independent because the same L$_X$--T  relation is used for clusters at all redshifts to compare their measured luminosity to the expected one. To distinguish the component of the selection bias that may mimic evolution of the L$_X$--T  relation, the effects of the local bias have to be taken into account. The local scaling relation bias leads to a division of cluster luminosities by a higher expectation value than the unbiased relation. To determine the evolution bias relative to this higher expectation value and not relative to the underlying cluster population, the bias curve has to be divided by the mean bias in the redshift range that was used to fit the "local" scaling relation. The bias curve was therefore rescaled by a factor of 0.8, corresponding to the red-dashed curve in Fig.\,\ref{fig:ltzbias}. To summarize, the rescaled bias curve shows the additional redshift-dependent bias of the logarithmic mean of the bolometric cluster luminosity divided by the luminosity resulting from the local L$_X$--T  relation relative to the already bias-affected local sample.

The bias curves of Fig.\,\ref{fig:ltzbias} have various characteristics that can easily be explained in terms of the underlying cluster sample. The non-rescaled bias is generally greater than 1 because for any flux-limited sample in the presence of scatter more clusters below the mean relation than above it are too faint to be included in the sample. With increasing redshift, the fraction of the cluster population that is excluded for being too faint increases, causing an increase in the fraction of clusters above the mean L$_X$--T  relation that have no counterpart below the relation. Hence, for a single flux limit the bias increases with redshift. This trend is visible in the bias curves for the three flux-limited subsamples, $f_\mathrm{min}\!=\!3*10^{-12} \mathrm{\ erg\ s^{-1}\ cm^{-2}}$(green), $f_\mathrm{min}\!=\!1*10^{-13} \mathrm{\ erg\ s^{-1}\ cm^{-2}}$(orange), and $f_\mathrm{min}\!=\!1*10^{-14} \mathrm{\ erg\ s^{-1}\ cm^{-2}}$(black). Up to a certain redshift threshold, the influence of selection bias on the subsamples is negligible because no significant part of the cluster population is excluded from the sample owing to the flux limit. Naturally, the unbiased redshift range increases with the survey sensitivity from about $z\!\sim\!0.05$ for the sample with $f_\mathrm{min}\!=\!3*10^{-12} \mathrm{\ erg\ s^{-1}\ cm^{-2}}$ to $z\!\sim\!0.2$ for the sample with $f_\mathrm{min}\!=\!1*10^{-13} \mathrm{\ erg\ s^{-1}\ cm^{-2}}$ and  $z\!\sim\!0.6$ if $f_\mathrm{min}\!=\!1*10^{-14} \mathrm{\ erg\ s^{-1}\ cm^{-2}}$. Surveys with a high limiting flux display a faster increase in bias with redshift than deeper surveys.

The characteristics of the mean bias curve for the combined cluster sample (blue curve in Fig.\,\ref{fig:ltzbias}) are affected by the dominant contribution in terms of cluster detections with increasing redshift shifting from the all-sky $f_\mathrm{min}\!=\!3*10^{-12} \mathrm{\ erg\ s^{-1}\ cm^{-2}}$-survey to the more sensitive but smaller solid-angle serendipitous surveys. The combined bias curve therefore decreases when the contribution from the $f_\mathrm{min}\!=\!1*10^{-13} \mathrm{\ erg\ s^{-1}\ cm^{-2}}$-survey becomes dominant at about $z\!\sim\!0.15$. At $z\!\sim\!0.8$, the curve drops again because from there toward higher redshift the contribution of the $f_\mathrm{min}\!=\!1*10^{-14} \mathrm{\ erg\ s^{-1}\ cm^{-2}}$-surveys dominates. From $z\!=\!0.8$ toward higher redshift, the bias increases again as outlined above.

The bias in the M--L$_X$ relation was not determined independently but based on the results of the L$_X$--T  relation. Cluster masses were set according to the local M--T relation  (Equ.\,\ref{mtloc}) and the ICM temperatures of the simulated cluster sample outlined before. The logarithmic mean of the cluster masses divided by the masses expected from the M--L$_X$ relation (Equ.\,\ref{mlloc}) was then calculated for the simulated sample to obtain a bias curve analogous to the one derived for the L$_X$--T  relation. By construction, the bias curves for the M--L$_X$ relation show the same features as those for the L$_X$--T  relation. However, the bias curve is inverted and the selection effects generally lead to an underestimation of the mean mass for a given luminosity.     

\begin{figure}[t]
	\flushleft
		\includegraphics[angle=-90,width=0.5\textwidth]{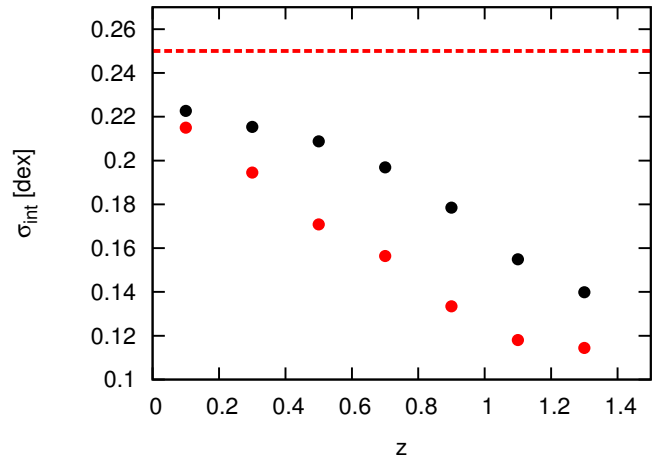}
	\caption{Redshift-dependent bias in the measured intrinsic scatter. Red-dashed line: true intrinsic scatter of the simulated samples. Black: fitted intrinsic scatter for $f_\mathrm{min}\!=\!1*10^{-14} \mathrm{\ erg\ s^{-1}\ cm^{-2}}$. Red: fitted intrinsic scatter for $f_\mathrm{min}\!=\!1*10^{-13} \mathrm{\ erg\ s^{-1}\ cm^{-2}}$.}
	\label{scbias}
\end{figure}

The intrinsic scatter in the cluster observables about the mean scaling relations is of great importance to the interpretation of results about scaling relation evolution. While the intrinsic scatter for local clusters is roughly known (see Pratt09), its redshift-dependence has not been well-constrained. Measuring the intrinsic scatter at high redshifts is challenging since in addition to the intrinsic scatter, the observed scatter in the evolution plots of  Sect.\ \ref{sec:ConstraintsFromTheCombinedClusterSample} has a number of other causes. First of all, measurement errors naturally have an influence on the observed scatter. Furthermore, increased scatter can also result from a change in the scaling-relation slope with redshift. Finally, selection effects of flux-limited cluster samples may have an influence on the observed scatter.

To estimate this bias in the L$_X$--T  relation, we measured the intrinsic scatter of various simulated cluster samples. The simulated cluster population has an intrinsic scatter in luminosity of 0.25 dex ($\sim\!60\%$). From this population, flux-limited samples with $f_\mathrm{min}\!=\!1*10^{-13} \mathrm{\ erg\ s^{-1}\ cm^{-2}}$ and $f_\mathrm{min}\!=\!1*10^{-14} \mathrm{\ erg\ s^{-1}\ cm^{-2}}$ were selected. The intrinsic scatter of the samples was then fitted in redshift bins centered around $z\!=\!0.1$, 0.3, 0.7, 0.9, 1.1, and 1.3. The results are shown in Fig\,\ref{scbias}.

For all subsamples, the fitting routine generally underestimates the real intrinsic scatter. The fitted scatter for local cluster samples is about 0.22 dex and comparable for the two samples. As more and more luminous clusters are excluded by the flux limit at higher redshifts, the fitted scatter shows a decreasing trend. As expected, the decrease is more rapid with redshift for the $f_\mathrm{min}\!=\!1*10^{-13} \mathrm{\ erg\ s^{-1}\ cm^{-2}}$-sample than the deeper $f_\mathrm{min}\!=\!1*10^{-14} \mathrm{\ erg\ s^{-1}\ cm^{-2}}$-sample. For the most distant subsamples at both flux limits ($z\!\sim\!1.3$), the estimated scatter is 0.12 and 0.14 dex, respectively, \ie \ the selection bias causes the scatter to be underestimated by about $50\%$.

\section{Results}
\label{sec:Results}

\subsection{Local scaling relations}
\label{sec:LocalScalingRelations}

The results for scaling relations fitted to observed cluster samples may differ significantly depending on the fitting scheme used. In this work, we use the BCES \label{bces} fitting method \citep{1996ApJ...470..706A} that has been widely used in recent studies and correctly accounts for intrinsic scatter about the mean relation and inhomogeneous measurement errors. However, several slightly different variations of this method exist. When choosing one of these alternatives, it is most important to distinguish between a fundamental independent and dependent variable. For the cosmological applications related to the results of this study, cluster mass is the fundamental property, and luminosity and temperature take the role of dependent variables. Therefore for the M--T and M--L$_X$ relation, the BCES(T$|$M) and BCES(L$|$M) method is used in this work. Owing to the large intrinsic scatter in the M--L$_X$ relation, the ICM temperature displays a tighter correlation with cluster mass than the X-ray luminosity. As a consequence, for the L$_X$--T  relation the BCES(L$|$T) scheme is used.

Only clusters with $z\!<\!0.3$ were considered for the local fit. This redshift threshold was chosen to be large enough to include a sufficiently large number of systems and improve the quality of our statistical analysis but small enough to keep evolution effects to a negligible level. To determine the influence of evolutionary effects, the fits were repeated with a maximum redshift of $z\!=\!0.2$ instead of $z\!=\!0.3$. The results derived with this lower redshift threshold are fully consistent with the sample at $z\!<\!0.3$, that is even though the expected evolution factor is non-negligible at $z\!=\!0.3$, no obvious evolutionary effects on the scaling fit for the entire local sample are observed. However, the statistical errors increase significantly with redshift because of the smaller sample size. The cluster properties were not rescaled by any assumed evolutionary model before the fit. 

The scaling relations derived from the combined local cluster sample are

\begin{equation}
M=(0.236\pm 0.031)*(T[\mathrm{keV}])^{1.76\pm 0.08} 10^{14} M_{\sun}\label{mtloc},
\end{equation}
\begin{equation}
L_X\!=\!(0.112\pm 0.031)*(T[\mathrm{keV}])^{2.53\pm 0.15} 10^{44} \mathrm{erg\ s^{-1}}\label{ltloc},
\end{equation}
\begin{equation}
M\!=\!(1.191\pm 0.104)*(L_X[10^{44} \mathrm{erg\ s^{-1}}])^{0.66\pm 0.04} 10^{14} M_{\sun},\label{mlloc}
\end{equation}

\noindent where $L_X$ is the bolometric X-ray luminosity and all properties are considered to be within $r_{500}$. However, for the analysis of scaling relation evolution with redshift, the relations derived by Pratt09 were considered instead of the fitted scaling relations because the former were derived from a more homogeneous sample with well-known selection criteria, that have smaller relative errors. The L$_X$--T  relation of Pratt09

\begin{equation}
L_X\!=\!(0.079\pm 0.008)*(T[\mathrm{keV}])^{2.70\pm 0.24} 10^{44} \mathrm{erg\ s^{-1}}\label{prattltloc}
\end{equation}

\noindent is consistent with our result in terms of both slope and normalization. No M--T relation is provided in Pratt09, but a fit to their sample with the BCES(T$|$M) method yields 

\begin{equation}
M=(0.291\pm 0.031)*(T[\mathrm{keV}])^{1.62\pm 0.08} 10^{14} M_{\sun},\label{prattmtloc}
\end{equation}

\noindent which is consistent with our result. Pratt09 provide an L-M instead of an M--L$_X$ relation. Fitting the inverse relation to their sample leads to 

\begin{equation}
M\!=\!(1.39\pm 0.07)*(L_X[10^{44} \mathrm{erg\ s^{-1}}])^{0.54\pm 0.03} 10^{14} M_{\sun},\label{prattmlloc}
\end{equation}

\noindent which differs slightly from our result at the $<\!2\sigma$ level.

The local scaling relations derived by means of the different fitting methods and the relations by Pratt09 are shown in Appendix\,\ref{sec:LocalScalingRelationsForTheCombinedClusterSample}.

\subsection{Evolution constraints from the combined cluster sample}
\label{sec:ConstraintsFromTheCombinedClusterSample}

The central goal of this work is to obtain a clearer understanding of the redshift evolution of X-ray scaling relations. The figures in this section help us to visualize these evolutionary trends. They show the redshift-dependent distribution of cluster properties divided by the expected value assuming local scaling relations. For the L$_X$--T  relation, this means that all cluster luminosities for instance are divided by $L_{z=0}(T)$, that is the luminosities inferred from the local L$_X$--T  relation at the measured cluster temperature, and that we plot the quantity $\frac{L_\mathrm{obs}}{L_{z=0}(T)}$ as a function of redshift.

Plotting the data this way, the properties of the local clusters have an approximately log-normal scatter around 1. A change in the normalization of the scaling relation with redshift translates into similar scatter around a different mean value in the evolution plot, whereas a change in the slope would result in a larger scatter around the mean value. We note, however, that this is not a very suitable test for changes in slope as a greater intrinsic scatter at earlier times also leads to larger scatter in the evolution plot and as outlined in Sect.\ref{sec:SelectionBiasEstimate}, selection biases may lead to an underestimation of the scatter of up to $50\%$. To investigate changes in the slope of high-z scaling laws, relations were fitted to the available $z\!>\!0.8$-clusters by means of the BCES method. Owing to the small sample size, however, the errors in the resulting relations are too large to allow independent constraints.

The self-similar model of cluster formation predicts the slopes of X-ray scaling relations to be redshift-independent and the normalization to vary in proportion to powers of $E(z)$ in the case of fixed overdensity. To test these predictions, a power-law $E(z)^\alpha$ was fitted to all cluster data points with $z\!>\!0.3$ for the M--T relation, \ie \ the unknown exponent $\alpha$ in

\begin{equation}
\frac{M_\mathrm{obs}}{M_{z=0}(T)}=E(z)^\alpha
\label{eq:evolfit}
\end{equation}

\noindent was constrained by fitting $\frac{M_\mathrm{obs}}{M_{z=0}(T)}$ versus $E(z)$ in log-log-space. In the redshift range $0.3\!<\!z\!<\!0.6$, an estimate of the influence of selection biases on the evolution results is challenging, since, in contrast to the more distant systems, the properties of these systems were obtained from both shallow surveys with a very high influence of selection biases but also deeper recent surveys. Therefore, this redshift range was excluded from the evolution fits for the L$_X$--T  and M--L$_X$ relation, for which selection biases play a more critical role than for the M--T relation. The excluded redshift range contains 45 sample clusters, while 65 clusters at $z\!>\!0.6$ are used for the evolution fit. Owing to the large number of luminous clusters from shallow surveys, whose lack of depth introduces the largest bias into the overall sample, an inclusion of this redshift range in the evolution fit would lead to a positive evolution result that does not trace the observed evolution for more distant clusters from deeper surveys. To account for the variations in the number density of clusters with redshift in our sample, \ie \ to avoid the result being exclusively determined by the large number of relatively low-redshift clusters with small errors, the data points were weighted by the inverse number of clusters in the corresponding redshift bin ($\Delta z\!=\!0.1$). Our evolution results are summarized in Table\,\ref{tab:sum}.\\

\begin{table}[t]
\caption{Evolution results based on the combined cluster sample. First column: Scaling relation. Second column: Observed evolution of scaling relations, bias effects have been accounted for by greater uncertainties. Third column: Evolution results including a tentative selection-bias correction. Fourth column: Self-similar expectations and self-similar predictions}\label{tab:sum}
\begin{center}
\begin{tabular}{cccc}
\hline
relation & observed evolution & bias-corrected & self-similar \\
\hline
M--T & $\propto E(z)^{-1.04\pm0.07}$ &  & $\propto E(z)^{-1}$ \\
L$_X$--T  & $\propto E(z)^{-0.23^{+0.12}_{-0.62}}$ & $E(z)^{-0.65\pm0.13}$ & $\propto E(z)^{+1}$ \\
M--L$_X$ & $\propto E(z)^{-0.93^{+0.62}_{-0.12}}$ & $E(z)^{-0.81\pm0.12}$ & $\propto E(z)^{-7/4}$ \\
\hline
\end{tabular}
\end{center}
\end{table}

\begin{figure*}[htb]
	\centering
		\makebox[0pt]{\includegraphics[angle=-90,width=1.0\textwidth]{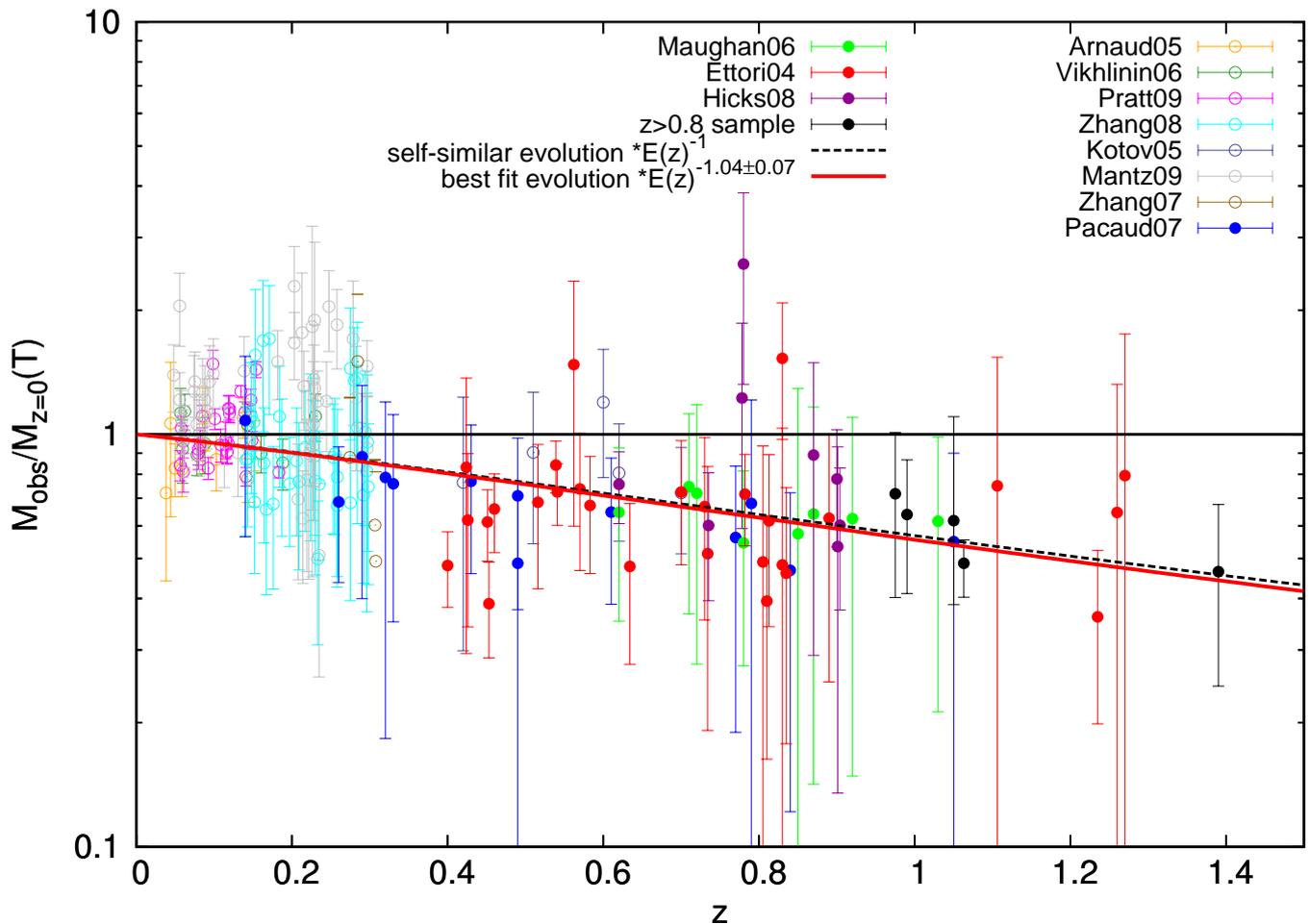}}
	\caption{Redshift evolution of the M--T relation. Black-dashed line: self-similar prediction ($\propto E(z)^{-1}$). Continuous red line: best-fit evolution ($\propto E(z)^{-1.04\pm0.07}$).}
	\label{fig:mtzdeltafix}
\end{figure*}

\noindent We first discuss the M--T relation, which is expected among all relations to most closely follow the self-similar predictions. Figure\,\ref{fig:mtzdeltafix} shows the redshift evolution of the M--T relation for the combined cluster sample. The best fit to the data corresponds to a redshift dependence of the normalization proportional to $E(z)^{-1.04\pm0.07}$, which is consistent with the self-similar prediction of $E(z)^{-1}$. We note that selection bias is not taken into account in this plot. However, for the the M--T relation this is not as important as for scaling relations including cluster luminosity. An analysis similar to the one shown in Fig.\,\ref{fig:mtzdeltafix} was performed with cluster radii defined with a variable density contrast $\Delta_z$ instead of a fixed $\Delta$ at cluster redshift. The results are consistent within the errors with those of Fig.\,\ref{fig:mtzdeltafix}, \ie \ using the redshift-dependent density contrast does not significantly influence our results about the evolution of the M--T relation. The M--T relation fitted to the $z\!>\!0.8$-clusters with sufficiently good X-ray data has a slope of $M\!\propto\!T^{1.59\pm 0.45}$, which is fully consistent with the local result. 

\begin{figure*}[htb]
	\centering
		\makebox[0pt]{\includegraphics[angle=-90,width=1.0\textwidth]{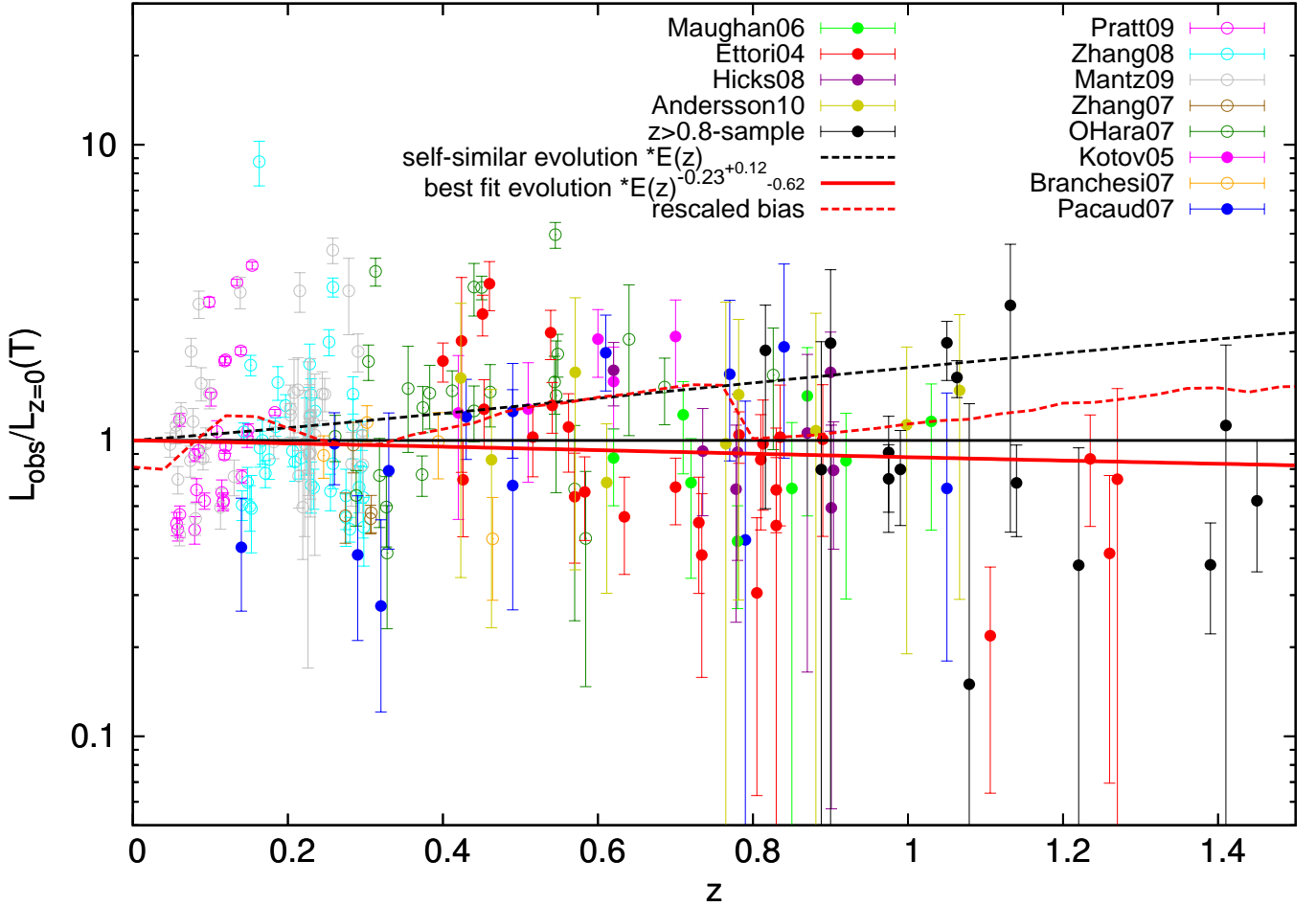}}
	\caption{Redshift evolution of the L$_X$--T  relation. Black-dashed line: self-similar prediction ($\propto E(z)$). Continuous red line: best-fit evolution  ($\propto E(z)^{-0.23^{+0.12}_{-0.62}}$). Red-dashed line: estimated mean bias for the combined sample rescaled to remove the effects of bias in the local scaling relation.}
	\label{fig:ltzdeltafix}
\end{figure*} 

\noindent Figure \ref{fig:ltzdeltafix} shows the redshift evolution of the L$_X$--T  relation for the combined cluster sample. The best-fit relation for the evolution is $E(z)^{-0.23\pm0.12}$, that is there is a slightly negative evolution. This result is clearly inconsistent with the self-similar prediction that the normalization increases with redshift in proportion to $E(z)^{+1}$. 

The evolution result was uncorrected for the estimated selection bias (see Sect.\ref{sec:SelectionBiasEstimate}) because this bias estimate relies on a toy model that is only approximately comparable to the real cluster sample. The error budget was instead increased by the estimated bias, \ie  the confidence region was enlarged by the size of the estimated bias (see Fig.\,\ref{fig:ltzbias}) in the direction of the supposed bias correction. This led to a final evolution result of $E(z)^{-0.23^{+0.12}_{-0.62}}$. As expected, applying an approximate bias correction based on the rescaled bias curve of Fig.\,\ref{fig:ltzdeltafix} before the fit as a test of the influence of selection biases results in an even more negative evolution result of $E(z)^{-0.65\pm0.13}$.

The slope of the L$_X$--T  relation fitted to the $z\!>\!0.8$-clusters with sufficiently accurate X-ray data available is $L\!\propto\!T^{3.12\pm 0.37}$. This result is slightly steeper but still consistent within the errors with the local slope derived by Pratt09. Owing to the small cluster sample and the large errors, this result heavily depends on the fitting method used and therefore provides no significant evidence of a steepening of the high redshift L$_X$--T  relation. As for the M--T relation, the use of a redshift-dependent density contrast $\Delta_z$ instead of a fixed $\Delta$ leads to comparable results with similar scatter about the mean relation.

\begin{figure*}[htb]
	\centering
		\makebox[0pt]{\includegraphics[angle=-90,width=1.0\textwidth]{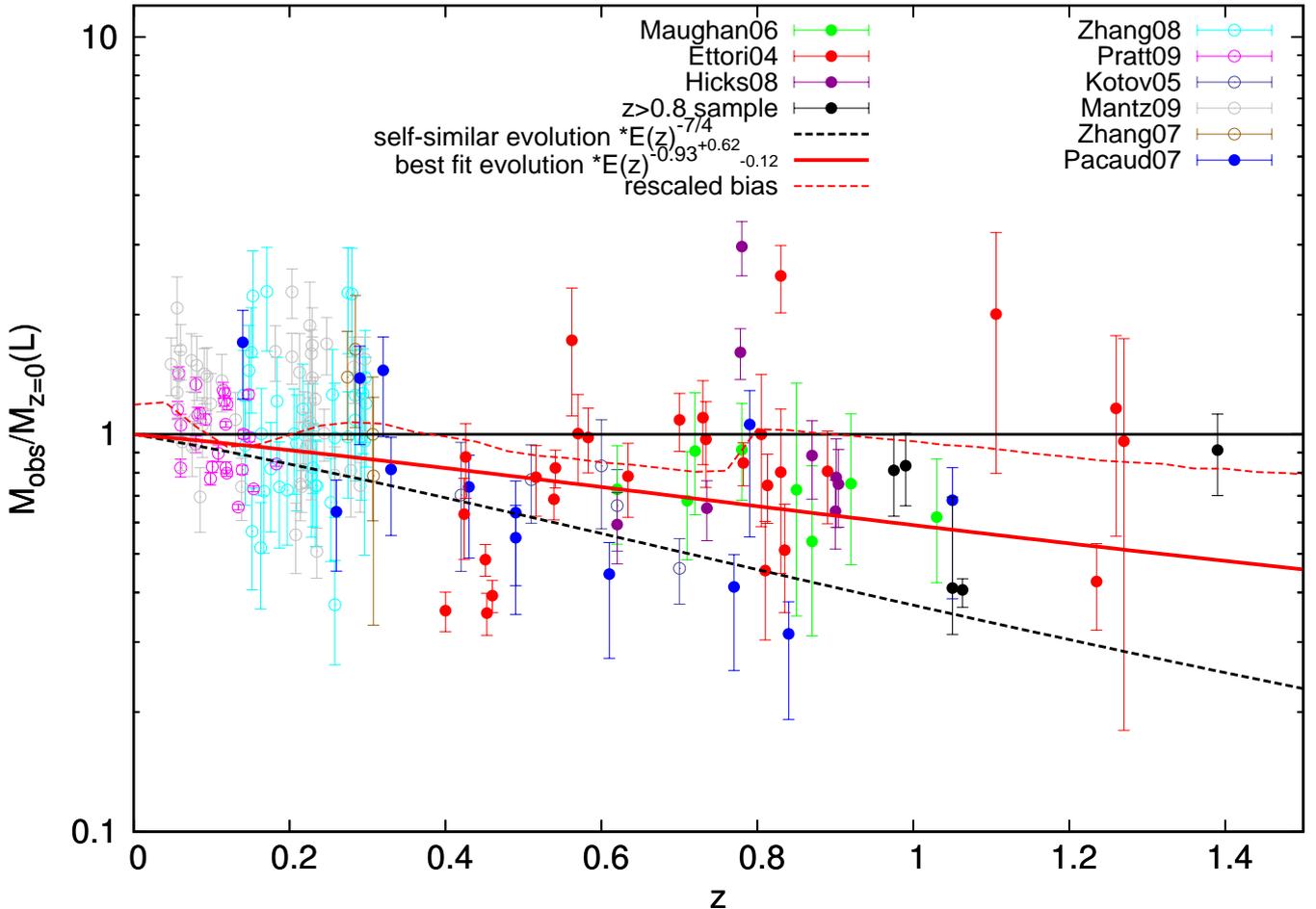}}
	\caption{Redshift evolution of the M--L$_X$ relation.  Black-dashed line: self-similar prediction ($\propto E(z)^{-7/4}$). Continuous red line: best-fit evolution ($\propto E(z)^{-0.93^{+0.62}_{-0.12}}$). Red-dashed line: estimated mean bias for the combined sample rescaled to remove the effects of bias in the local scaling relation.}
	\label{mlz}
\end{figure*}

\noindent Figure \ref{mlz} shows the redshift evolution of the M--L$_X$ relation for the combined cluster sample. The best-fit relation is $E(z)^{-0.93\pm0.12}$, \ie \ negative evolution as predicted by the self-similar model. However, our result is significantly less steep than the self-similar prediction of $E(z)^{-7/4}$. As for the M--T relation, the estimated selection bias is taken into account in the error budget and leads to the evolution being proportional to $\propto E(z)^{-0.93^{+0.62}_{-0.12}}$. This observed evolutionary trend is close to the one expected after combining the results for the evolution of the L$_X$--T  and M--T relations, which would be $\propto E(z)^{-0.90}$. Applying an approximate bias correction to test the influence of selection biases as for the L$_X$--T  relation results in a slightly less negative evolution fit of $E(z)^{-0.81\pm0.12}$. In Pratt09, a bias-corrected local L$_X$--M  relation is provided. Using the inverted bias-corrected BCES(L$|$M)-relation ($M\!=\!(1.64\pm 0.07)*(L_X[10^{44} \mathrm{erg\ s^{-1}}])^{0.52\pm 0.03}$) and correcting for the total estimated evolution bias (not the curve rescaled to remove the effects of bias in the local scaling relation) leads to an evolution result of $E(z)^{-0.90^{+0.35}_{-0.15}}$. The estimated errors in this result include statistical errors and an estimate of the systematical error caused by an inexact bias correction.

The slope of the fitted high redshift M--L$_X$ relation is $M\!\propto\!L^{0.70\pm 0.21}$, which is consistent with the local result. Using the $\Delta_z$-scheme instead of a redshift-independent density contrast again leads to similar results. The observed scatter in cluster properties about the mean relation for the high redshift clusters is consistent with the local scatter in all three relations. The real scatter about the L$_X$--T  and M--L$_X$ relation for distant clusters may be up to a factor of two larger than the observed result because of the influence of selection biases (see Sect.\,\ref{sec:SelectionBiasEstimate}). However, owing to conservative error estimates, no constraints on the intrinsic scatter in cluster properties can be placed based on the measured total scatter for the distant cluster sample.

\section{Discussion}
\label{sec:Discussion}

\subsection{Stability of results}
\label{sec:StabilityOfResults}

The results on scaling relation evolution presented in the previous section were obtained by means of a number of input assumptions and results of preceding studies that have an influence on the obtained results and may introduce additional errors. Throughout the remainder of this section, we briefly discuss the stability of the results under these assumptions.

The assumed local scaling relations have a direct influence on the observed evolution. For our analysis in Sect.\,\ref{sec:ConstraintsFromTheCombinedClusterSample}, the redshift evolution of scaling relations was determined using the local scaling relations of Pratt09. Using the relations derived for the entire local combined sample (see Sect.\ref{sec:LocalScalingRelations}) instead of these results does not lead to fundamentally different findings on the evolution coefficient. For the M--T relation, using the relation derived from the combined sample leads to a best-fit evolution of $E(z)^{-1.15\pm 0.06}$, which is only slightly different from the result for the Pratt09 relation ($E(z)^{-1.04\pm0.07}$). Using the L$_X$--T  relation fitted to our sample implies an evolution result of $E(z)^{-0.36\pm 0.12}$ instead of $E(z)^{-0.23^{+0.12}_{-0.62}}$ and for the M--L$_X$ relation $E(z)^{-1.06\pm 0.13}$ instead of $E(z)^{-0.93^{+0.62}_{-0.12}}$, both of which are fully consistent with the results presented above for both relations.
  
An incomplete or incorrect homogenization scheme applied to the different subsamples naturally influences the evolution results. However, the combined cluster sample provides no hints that this might be a major problem (see Appendix\,\ref{sec:ComparisonOfClusterPropertiesForSystemsIncludedInMoreThanOneSubsample}). Incorrect homogenization would be visible as larger scatter about the mean behavior in the cluster sample or different evolutionary trends for different subsamples. Taking into account selection biases, these significant trends are not observed for the cluster sample (see Sect.\ref{sec:ComparisonToPublishedResults2}).

Our study again highlights the importance of selection biases when investigating scaling relation evolution and the problems inherent to small cluster samples over a limited redshift range. Although the simulated cluster sample used to estimate bias effects in this study is only a rough approximation of the true situation, it reveals the apparent evolutionary trends caused by selection effects, which have been taken into account in the estimated errors. Despite the lack of knowledge about the exact effects of selection bias in a highly inhomogeneous combined cluster sample such as ours, at least a fair estimate of the influence on the evolution results can be given. For both the M--L$_X$ and the L$_X$--T  relations, a bias correction of the evolution results would render the difference to the self-similar predictions even more significant. Our finding about the inconsistency with the self-similar model can therefore not be attributed to selection effects.

The X-ray properties of the systems within a cluster sample and their evolution with redshift might depend on the cluster selection strategy. While the subsamples of clusters selected by means of the Sunyaev-Zel'dovich effect (SZ) and their optical/infrared (IR) properties are not sufficiently extensive to place independent constraints on scaling relation evolution, the X-ray, SZ,and optical/IR-selected subsamples display no obvious differences in their evolutionary trends.
 
In recent studies, we note that different definitions of the density contrast $\Delta$ and the resulting cluster radii have been used. However, the choice of either a fixed $\Delta$ at the cluster redshift or a redshift-dependent $\Delta_z$ has no significant influence on the determined evolution results.

\subsection{Comparison to published results}
\label{sec:ComparisonToPublishedResults2}

We now briefly discuss previously published results on scaling relation evolution comparing these to our findings:
\begin{description}
\item[Kotov05: ]For the M--T relation, their evolution result of $\propto E(z)^{-0.88\pm 0.23}$ is consistent with both our results and the self-similar expectation. For the L$_X$--T  relation, Kotov05 find the normalization to be $\propto (1+z)^{1.8\pm 0.3}$, \ie \ a positive evolution that is even steeper than self-similar. This trend is easily visible for their sample in Fig.\,\ref{fig:ltzdeltafix}. The seven clusters of the Kotov05 sample cover a redshift range of $0.4\!<\!z\!<\!0.7$. Comparing this subsample with the remainder of the combined sample in this redshift range reveals that instead of describing the true evolution, this result can be attributed to selection effects, in addition to both limited sample size and redshift range. We note that the evolution result derived in Kotov05 approximately traces the bias curve for the sample with $f_\mathrm{min}\!=\!1*10^{-13} \mathrm{\ erg\ s^{-1}\ cm^{-2}}$ visible in Fig.\,\ref{fig:ltzbias}. Although no uniform flux limit can be assigned to this archival sample, $f_\mathrm{min}\!=\!1*10^{-13} \mathrm{\ erg\ s^{-1}\ cm^{-2}}$ represents an appropriate estimate of the mean flux limit,  \ie \ the bias estimate suggests that after correcting for selection effects the observed evolution should be close to zero.   
\item[Maughan06: ]Their result for the L$_X$--T  evolution is $\propto (1+z)^{0.8\pm 0.4}$ when using the local relation of \citet{1999MNRAS.305..631A} and $\propto (1+z)^{0.7\pm 0.4}$ for the \citet{1998ApJ...504...27M} relation, \ie \ a slightly positive evolution. Using the scaling relation derived by Pratt09 instead removes this evolutionary trend, causing the evolution of the Maughan06 subsample ($\propto E(z)^{-0.34\pm 0.31}$) to be consistent with the results of our study.
\item[OHara07: ]Their result about the evolution of the core-included $L_X$-T relation within $r_{500}$ is $E(z)^1(1+z)^{-1.90^{+1.17}_{-1.11}}$, \ie \ consistent within the errors with our result.
\item[Branchesi07: ] No significant evolution of the L$_X$--T relation in the redshift range $0.3\!<\!z\!<\!1.3$ was observed. However, the normalization for this redshift range was found to be about a factor of $\sim 2$ higher than suggested by the local relations of \citet{1998ApJ...504...27M} and \citet{1999MNRAS.305..631A}. The differences from the results derived in this work can mostly be attributed to the use of a different local scaling relation.
\item[Pacaud07: ]Since a detailed selection function was derived in this work, this analysis provides an important insight into the influence of selection effects. Before correcting for those, their result about the evolution of the L$_X$--T  relation is roughly similar to that of Kotov05. Afterwards,  they obtain an evolution factor of $\propto E(z)^1(1+z)^{-0.07^{+0.41}_{-0.55}}$, \ie \ slightly less than self-similar. This result is still marginally inconsistent with ours. We note, however, that a significant part of their sample was not included here because the temperatures were below 2 keV.
\item[Ettori04: ]Their inferred evolution of the L$_X$--T  relation varies drastically depending on whether clusters below $z\!<\!0.6$ are included in the fit. Using the local relation of \citet{1998ApJ...504...27M}, they find the evolution to be $\propto (1+z)^{0.62\pm 0.28}$ for the entire sample and $\propto (1+z)^{0.04\pm 0.33}$ if only clusters with $z\!>\!0.6$ are considered. Using the relation of Pratt09 leads to results consistent with those from the entire combined sample at least for the $z\!>\!0.6$-clusters ($\propto E(z)^{-0.60\pm 0.19}$). The marked difference between the clusters at $z\!<\!0.6$ and the more distant systems can probably be attributed to selection effects, that is to the fact that the low-z sample consisting mostly of archival clusters detected in relatively shallow observations, as discussed above for the results of Kotov05, in accordance with which Ettori04 find the M--T evolution to be consistent with the self-similar prediction.
\item[\citet{2009ApJ...692.1033V}: ] When applying X-ray galaxy cluster data to cosmological tests these authors also provide a new evaluation of the L$_X$--M relation and its evolution with redshift in their Equ.\, 22.
In contrast to our calculations, this relation is determined for
luminosities in the 0.5 - 2 keV band. Using the results of Pratt
about the difference between the  bolometric and 0.5 - 2 keV band
scaling relations for comparison, we find that the results
of Vikhlinin et al. and ours are in very good agreement for the
zero redshift relation. Analyzing the redshift-dependent term
in the relation inferring L$_X$ from M, we find a term of
$ E(z)^{1.85 \pm 0.42}$ for Vikhlinin et al. and a term of
$ E(z)^{1.73^{+0.41}_{-0.73}}$ by inverting our Equ.\,\ref{eq:mlbiascorr}.
There is again good agreement within the large error bars.
Taking the mean trends of both relations, we find a difference
in the evolution parameter of $\sim 5\%$ at redshift $z\!=\!0.8$
and  $\sim 10\%$  at $z\!=\!1.5$. Therefore, the difference between the
two relations is smaller than can be detected with any current
data set.
\item[\citet{2010ApJ...709...97L}: ]In their study, the $\mathrm{M}_{200}$--L$_X$ relation for 206 X-ray-selected galaxy groups from the COSMOS survey is determined in the redshift range $0.22\!<\!z\!<\!0.9$ by means of stacked weak-lensing mass estimates. The derived evolution of $\propto E(z)^{-1.77\pm 0.9}$ is consistent with our result.
\end{description}

\noindent In summary, the present study provides a clearer picture of the scaling-relation
evolution in three aspects. The first is that the use of the very early results by \citet{1998ApJ...504...27M}
and \citet{1999MNRAS.305..631A} as a local reference of the scaling relations
introduces a positive evolutionary trend as already shown in Branchesi07. Using more recent results with higher quality statistics, and in particular
the use of the weakly biased (not flux-limited) REXCESS sample results by
Pratt09 removes most of this trend. The second aspect is that the overview of a larger set of
cluster samples with different biases leads to the identification of bias effects from high flux-limits. Thirdly, the extension of the redshift range to newly detected high-redshift clusters
increases the leverage for the evaluation of evolutionary effects.

\section{Implications of results}
\label{sec:ImplicationsOfResults}

\subsection{Implications for the thermal history of the ICM}
\label{sec:ImplicationsForTheThermalHistoryOfTheIcm}

The observed modification of the evolution of scaling relations with respect to the self-similar model is caused by changes in the thermodynamical state of the ICM. Therefore, different evolutionary trends are the signature of different histories of heating and cooling processes in the clusters. Comparing observational results with the predictions made by hydrodynamical cosmological simulations permits us to assess whether the heating scheme implemented in the simulations is realistic. A variety of existing simulations taking into account different sources of heating and assumptions on the time evolution of non-gravitational heating and cooling allow a constraint of the most realistic heating scheme.

Previous attempts these comparison analyses were complicated by the inconsistent observational results about scaling relation evolution. Owing to updated local scaling relations based on morphologically unbiased cluster samples and the availability of a larger cluster sample over a wider redshift range, evolution constraints derived from the combined cluster sample used in this work permit a meaningful comparison. The simulations considered on that account are the Millennium Gas Simulations (MGS, \citet{2010ApJ...715.1508S}, \citet{2010MNRAS.408.2213S}), a series of hydrodynamical resimulations of the original dark-matter-only Millennium simulations \citep{2005Natur.435..629S}. The MGS runs incorporate a $(500 h^{-1}\ \mathrm{Mpc})^3$-volume with the same initial conditions and cosmological model as the original Millennium run, but include an equal number of gas particles in addition to the $5*10^{8}$ dark matter particles.

The MGS consist of three simulation runs with different implementations of heating schemes:
\begin{description}
\item[GO run:]This simulation does not include any additional non-gravitational heating sources. As a consequence, it mostly reproduces the self-similar model expectations but drastically fails to reproduce the observed local galaxy cluster X-ray scaling relations, hence is not considered in the comparison analysis.
\item[PC run:]This run employs a simplistic preheating model, raising the entropy of the gas particles to $200 \mathrm{\ keV \ cm^2}$ at $z\!=\!4$. In addition to preheating, radiative cooling according to the cooling function of \citet{1993ApJS...88..253S} is implemented.
\item[FO run:]The third simulation run includes no radiative cooling but incorporates a model of the energy input by supernova and AGN feedback computed by means of a semi-analytic model of galaxy formation, \ie \ a gradual injection of energy into the surrounding ICM gas.
\end{description}

The PC and FO runs represent two opposing heating schemes, one assuming an early preheating of the gas before its accretion onto the cluster, the other incorporating non-gravitational heating as a relatively recent and ongoing process. Both runs are able to reproduce the observed local scaling relations, \ie \ they predict the correct properties of the ICM for low-redshift clusters. However, owing to the contrasting heating schemes, the two  models make opposing predictions about the thermal evolution of the ICM with redshift. Table\,\ref{sim} gives an overview of the MGS evolution results fitted in the redshift range $0\!<\!z\!<\!1.5$.

\begin{table}[htbp]    
\begin{center}

\caption{Results on the redshift evolution of X-ray scaling relations from the Millennium Gas Simulations \citep{2010MNRAS.408.2213S}.}\label{sim}
\begin{tabular}{cc}
\hline
{\bf GO run}   & {\bf redshift evolution} \\
\hline
M--T &  $\propto E(z)^{-1}*(1+z)^{0.590\pm 0.023}$\\
L$_X$--T  & $\propto E(z)^{1}*(1+z)^{0.37\pm 0.06}$ \\
M--L$_X$ &  $\propto E(z)^{-7/4}*(1+z)^{0.160\pm 0.036}$\\
\hline
{\bf PC run}   & {\bf redshift evolution} \\
\hline
M--T &  $\propto E(z)^{-1}*(1+z)^{-0.074\pm 0.011}$\\
L$_X$--T  &  $\propto E(z)^{1}*(1+z)^{-1.77\pm 0.16}$ \\
M--L$_X$ &  $\propto E(z)^{-7/4}*(1+z)^{0.594\pm 0.073}$\\
\hline
{\bf FO run} & {\bf redshift evolution} \\
\hline
M--T &  $\propto E(z)^{-1}*(1+z)^{0.438\pm 0.025}$\\
L$_X$--T  &  $\propto E(z)^{1}*(1+z)^{0.76\pm 0.05}$ \\
M--L$_X$ &  $\propto E(z)^{-7/4}*(1+z)^{-0.494\pm 0.010}$\\
\hline
\end{tabular}
\end{center}
\end{table}

\begin{figure}[t]
	\flushleft
		\includegraphics[angle=-90,width=0.5\textwidth]{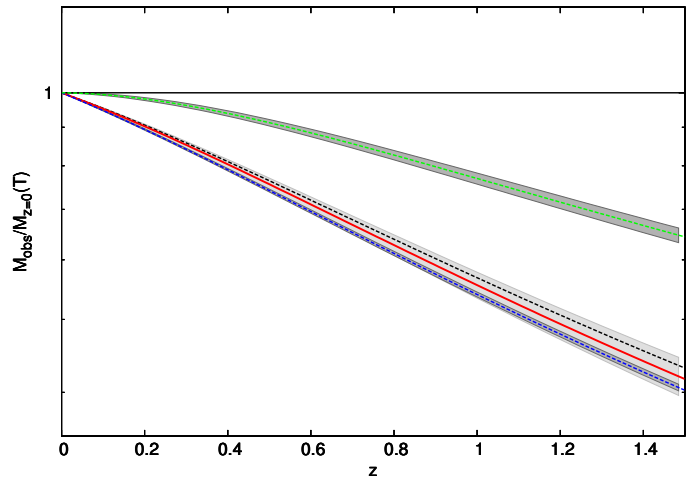}
	\caption{Redshift evolution of the M--T relation. Continuous red line and light grey confidence area: observed evolution. Green-dashed line and dark grey confidence area: MGS FO run. Blue-dashed line and dark grey confidence area: MGS PC run. Black-dashed line: self-similar prediction.}
	\label{mtzsim}
\end{figure}

\begin{figure}[t]
	\flushleft
		\includegraphics[angle=-90,width=0.5\textwidth]{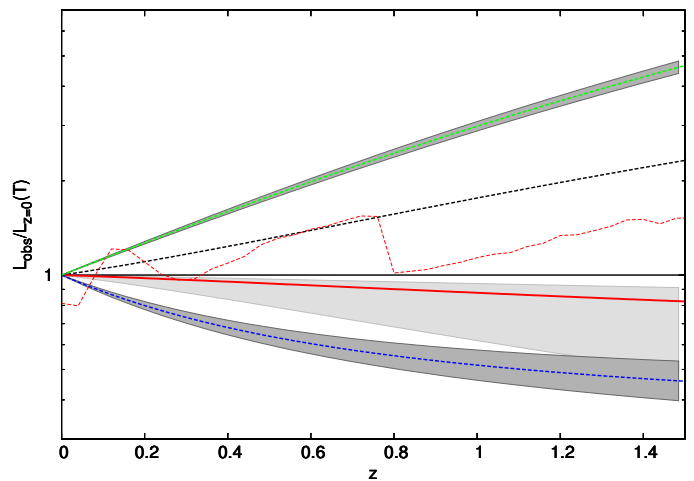}
	\caption{Redshift evolution of the L$_X$--T  relation. Red-dashed line: Estimated mean bias of the combined sample rescaled to remove the effects of bias in the local scaling relation. Additional lines and confidence areas have the same meaning as in Fig.\,\ref{mtzsim}.}
	\label{ltzsim}
\end{figure}

\begin{figure}[ht]
	\flushleft
		\includegraphics[angle=-90,width=0.5\textwidth]{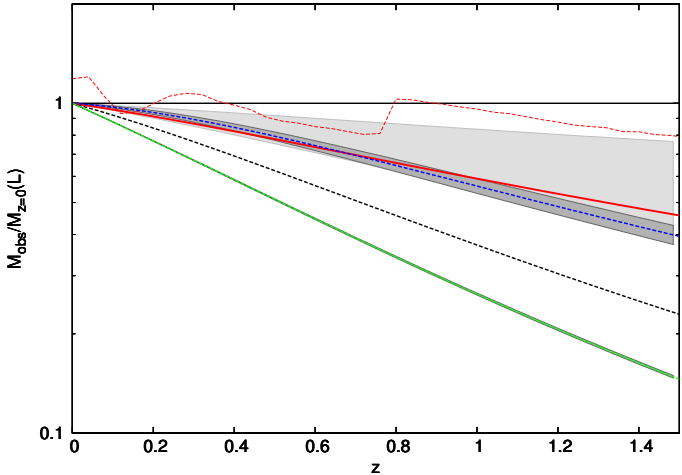}
	\caption{Redshift evolution of the M--L$_X$ relation. Lines and confidence areas have the same meaning as in Fig.\,\ref{ltzsim}.}
	\label{mlzsim}
\end{figure}

\noindent Depending on the selection criteria of the cluster sample, the adopted fitting scheme, and since the fitted slope and normalization of scaling relations are not independent of each other, the measured normalization of local scaling relations differs by up to a factor of two (see \eg \ the results of \citet{2005AA...441..893A} and \citet{2009AA...498..361P}). Therefore, instead of directly comparing the redshift-dependent normalization derived from the MGS to the results of this work, we do not take into account the different local normalizations but only the evolution with respect to the local values. Fig.\,\ref{mtzsim}, \ref{ltzsim}, and \ref{mlzsim} show the redshift evolution of the normalization of the M--T, L--T, and M--L$_X$ relations with respect to its local value for both the MGS PC and FO runs and the observational results deduced in this work. We note that in contrast to this work, \citet{2010MNRAS.408.2213S} assumed the self-similar evolution model and then fitted the observed difference from this model as powers of  $(1+z)$. 

It is clearly visible that the FO simulation provides no good description of the observed evolution for all three relations. In contrast, the PC run shows good agreement with the observations. For the M--T and M--L$_X$ relations, the PC results are fully consistent with ours within the errors. The prediction for the L$_X$--T  relation shows slight deviations from our result at the $2\sigma$ level for $z\!\lesssim\!1.2$. This difference can partly be attributed to the different functional form assumed for the evolution fit in \citet{2010MNRAS.408.2213S}. 

Care has to be taken when interpreting the constraints on the ICM thermal history deduced from this finding because none of the MGS runs provide a complete model of the necessary ICM heating and cooling processes. As an example, the PC run does not include ongoing heating, which is known to be of crucial importance to balance the cooling in cluster cores. Such an addition of mild ongoing feedback to the model is likely to bring the predicted evolution for the L$_X$--T  relation in even closer agreement with the observed trend (see Fig.\,\ref{ltzsim}). Nevertheless, the observed evolution of X-ray scaling relations provides strong evidence of the preheating scenario. A further refinement of the simulations, \eg \ the combination of preheating with mild ongoing feedback, is desirable to permit more detailed comparisons between the observed and the simulated evolutions. 

\subsection{Implications for the eROSITA cluster survey} 
\label{sec:ImplicationsForTheErositaClusterSurvey}

eROSITA \label{erosita} is the main instrument on the Spektrum-Röntgen-Gamma mission scheduled for launch in 2012 and will carry out a new all-sky X-ray survey in the energy range from 0.1 to 10 keV (see \citet{2010SPIE.7732E..23P} and \citet{2007SPIE.6686E..36P}). One of the main science goals of the project is to provide a sample of $\sim\!100\,000$ X-ray selected galaxy clusters. A cluster catalogue of this size is necessary to test cosmological models to higher accuracy and place reliable constraints on cosmological parameters such as the dark energy equation of state \citep{2005astro.ph..7013H}.

The achievable accuracy of the parameter constraints derived from the eROSITA all-sky survey depends on the number of cluster detections and the knowledge of X-ray scaling relations up to high redshift. Whether a cluster can be detected by eROSITA is mainly determined by the cluster's soft band X-ray luminosity and its distance. Since the mean cluster luminosity at a given redshift depends on the evolution of X-ray scaling relations, the results of this work have a direct influence on the expected number of eROSITA clusters to be detected at high redshift and therefore on the expected cosmological constraining power of the mission.    

The expected redshift distribution and total number of eROSITA clusters was recalculated by means of the cluster counter presented in chapter 8 of \citet{2010PhDT........M}. The cluster counter provides an estimate of the number of eROSITA cluster detections based on various simplifying assumptions. The assumed criterion for a cluster detection is whether the number of photons detected from the observed system exceeds the count limit $c_\mathrm{lim}$. Until now, the eROSITA count limit has not been known in detail owing to its dependence on as of now unspecified instrumental parameters such as the eROSITA point spread function, and furthermore its significant dependence on the total X-ray background, which varies with sky position, exposure time and other conditions. A constant detection count limit of $c_\mathrm{lim}\!=\!100$ is currently used as a conservative estimate.
  
The number of clusters in each redshift and luminosity bin was set according to the luminosity function used in Sect.\,\ref{sec:SelectionBiasEstimate}. This luminosity function is based on a cluster mass function resulting from a cosmological model consistent with the seven-year WMAP results \citep{2010arXiv1001.4538K} and was converted into a luminosity function by means of the L-M relation and the evolution results of Sect.\,\ref{sec:ConstraintsFromTheCombinedClusterSample}. Further necessary input data for the cluster counter are an all-sky exposure map for the special geometry of the eROSITA survey and an all-sky map of Galactic neutral hydrogen. The ICM temperature was set according to the (ROSAT band) L$_X$--T  relation of Pratt09 and the observed evolution of Sect.\,\ref{sec:ConstraintsFromTheCombinedClusterSample}. 

Thereafter, the eROSITA count rate at each position in the sky, cluster luminosity, and redshift is calculated by means of XSPEC, assuming an absorbed Mekal model with the parameters z, $L_X$, T, and $n_\mathrm{H}$. The count rate is then converted into a number of detected photons by multiplying it with the exposure value at the respective sky position. If the number of detected photons for a given set of parameters exceeds $c_\mathrm{lim}$, the number of clusters determined by the luminosity function for the given redshift and luminosity is added to the cluster number in the current redshift bin and sky position. The resulting redshift distribution and total number of detected clusters is presented in the following section for three L$_X$--T  evolution models: The self-similar model, \ie \ positive evolution, the no-evolution scenario assumed in \citet{2010PhDT........M}, and the slightly negative evolution found in this work.  

Table\,\ref{nerosita} shows the total number of expected cluster detections with eROSITA under the assumption of a count limit $c_\mathrm{lim}\!=\!100$. In addition to the all-sky expectations, the number of "extragalactic" detections refers to clusters with Galactic latitude $|b|\!>\! 20^\circ$. The total number of achievable cluster detections is expected to be closer to this last number, since the high column density of the absorbing Galactic interstellar medium and high density of other X-ray and stellar sources in the Galactic plane make cluster detections in this area challenging. For the extragalactic area, the number of expected clusters with $z\!>\!0.8, 1, 1.2, 1.4$, and 1.6 is also listed.

\begin{table}[t]    
\begin{center}
\caption{Number of expected cluster detections with eROSITA assuming self-similar, no, or slightly negative evolution of the L$_X$--T  relation. The rows labeled "extragalactic" refer to a Galactic latitude $|b|\!>\! 20^\circ$. A count limit of $c_\mathrm{lim}=100$ was assumed.}\label{nerosita}
\begin{tabular}{ccccccc}
\hline

{\bf self-sim.}   & {\bf $N_\mathrm{tot}$}& {\bf $N_{z>0.8}$}& {\bf $N_{z>1}$}& {\bf $N_{z>1.2}$}& {\bf $N_{z>1.4}$}& {\bf $N_{z>1.6}$} \\
\hline
all sky & 132\,787 &&&&& \\
extragal. & 97\,195 & 5\,834 & 2\,062 & 699 & 227 & 69 \\
\hline
{\bf no evol.}   & {\bf $N_\mathrm{tot}$}& {\bf $N_{z>0.8}$}& {\bf $N_{z>1}$}& {\bf $N_{z>1.2}$}& {\bf $N_{z>1.4}$}& {\bf $N_{z>1.6}$} \\
\hline
all sky & 120\,965
 &&&& \\
extragal. & 88\,238 & 4\,297 & 1\,414 & 447 & 135 & 38 \\
\hline
{\bf best fit}   & {\bf $N_\mathrm{tot}$}& {\bf $N_{z>0.8}$}& {\bf $N_{z>1}$}& {\bf $N_{z>1.2}$}& {\bf $N_{z>1.4}$}& {\bf $N_{z>1.6}$} \\
\hline
all sky & 114\,803 &&&& \\
extragal. & 83\,603 & 3\,505 & 1\,083 & 320 & 90 & 22 \\
\hline
\end{tabular}
\end{center}
\end{table}   

\begin{figure}[t]
	\flushleft
	\vspace{1cm}
		\includegraphics[width=0.5\textwidth]{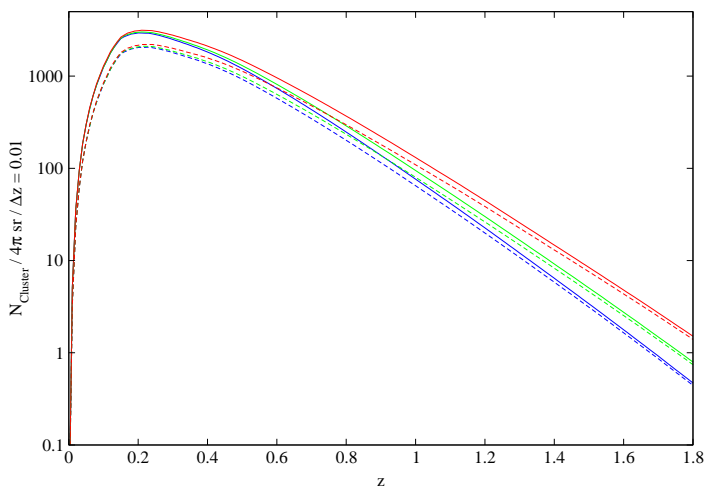}
	\caption{Achievable number of cluster detections with redshift within the redshift interval $[z-0.005;\ z+0.005]$ for eROSITA. Continuous lines: All sky. Dashed lines: Extragalactic ($|b|\!>\!20^\circ$). Black: Self-similar L$_X$--T  evolution ($\propto E(z)^{+1}$). Green: No L$_X$--T  evolution. Red: Best fit L$_X$--T  evolution}
	\label{fig:nz}
\end{figure}

\begin{figure}[ht]
	\flushleft
		\includegraphics[width=0.5\textwidth]{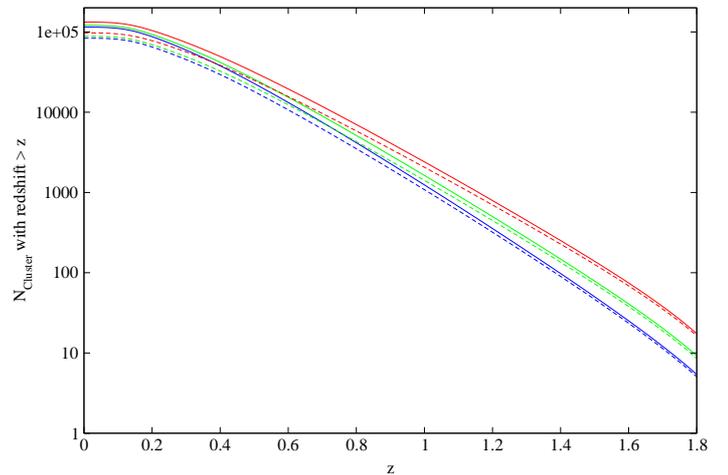}
	\caption{Achievable number of cluster detections with redshift $>\!z$ for eROSITA. The lines have the same meaning as in Fig.\,\ref{fig:nz}.}
	\label{fig:nzsum}
\end{figure}

\noindent As can be seen in Fig.\,\ref{fig:nz}, which shows the number of cluster detections with redshift within the redshift interval $[z-0.005;z+0.005]$ and Fig.\,\ref{fig:nzsum}, the achievable number of cluster detections with redshift $>\!z$, a change in the assumed evolution model has a significant influence on the number of eROSITA cluster detections. However, the total number of clusters that should be detected is hardly affected because the majority of the eROSITA clusters are expected to be discovered at low redshifts, where the scaling relation evolution is of little importance. As an example, the total number of clusters for the extragalactic area is decreased by $\sim\!5\%$ when changing from the previously assumed no-evolution scenario to the negative evolution found in this work. 

The situation is clearly different for distant cluster detections. Owing to the smaller number of luminous high-redshift clusters in the negative evolution scenario, the number of expected detections is smaller by $\sim\!18\%$ at $z\!=\!0.8$ and by $\sim\!28\%$ at $z\!=\!1.2$ relative to the no-evolution scenario. However, even with these corrected cluster expectations, eROSITA is still likely to increase the sample of known $z\!>\!0.8$-clusters by about a factor of 50 from present numbers.

\section{Summary and conclusions}
\label{sec:SummaryAndConclusions}

The main goal of this study has been to to investigate the redshift evolution of galaxy cluster X-ray scaling relations by means of a combined cluster sample. To this end, a cluster sample was compiled from both recent publications and newly discovered clusters provided by the XMM-Newton Distant Cluster Project (XDCP). Our gathered sample of recently discovered distant clusters has allowed tighter constraints to be made on scaling relation evolution than possible in previous studies.

The definition of cluster observables slightly differs between the individual publications used to compile the combined cluster sample. A homogenization scheme that accounts for these different analysis schemes was therefore applied to the subsamples. In detail, the definition of cluster radii, either relying on a fixed density contrast $\Delta$ at the cluster redshift or a redshift-dependent average overdensity $\Delta_z$ and the different values for the density contrast were corrected for. Furthermore, the applied corrections take into account the slight differences in the assumed cosmological parameters, in the energy band for which $L_X$ is given.

On the basis of data for clusters at $z\!<\!0.3$ included in the combined sample,  local M--T, L$_X$--T, and M--L-relations were fitted. The derived relations generally agree well with the published results of Pratt09 and show deviations from the self-similar model similar to those found in earlier studies, \eg \ a steeper L$_X$--T  relation. 

Typical X-ray selected cluster samples can be approximated to be flux-limited. Various selection biases complicate the analysis of these flux-limited samples and are important to the interpretation of results on scaling relations and their evolution. The bias inherent to the sample used in this work was estimated by means of comparable cluster samples selected from a simulated cluster population. In detail, selection biases were found to raise the measured normalization of the local L$_X$--T  relation and decrease the apparent scatter about the mean relation for high-z clusters. For the L$_X$--T  relation, the bias appears to generate a positive evolution, whereas for the M--L$_X$ relation it generates the opposite. 
  
The redshift evolution of X-ray scaling relations was investigated in the redshift range $0\!<\!z\!<\!1.46$. Throughout this redshift range, no significant variation in the slope of the relations was found. The normalization, however, evolves with redshift for all three examined relations. For the M--T relation, the measured evolution is  $\propto\!E(z)^{-1.04\pm0.07}$, consistent with the self-similar prediction. The results for the L$_X$--T  ($\propto\! E(z)^{-0.23^{+0.12}_{-0.62}}$) and M--L$_X$ relation ($\propto\! E(z)^{-0.93^{+0.62}_{-0.12}}$), however, differ significantly from the model predictions. As for the local relations, these deviations indicate that the influence of non-gravitational ICM heating and cooling is not negligible. The inconsistent results of recent studies have been found to be caused by limited sample sizes and redshift ranges in combination with selection biases and to a lesser degree by the use of different local relations. 

On the basis of our results and the local scaling of Pratt09 and assuming that $h\!=\!0.70$, the M--T, L--T, and M--L$_X$ relations for cluster properties within $r_{500}$ have the explicit form
\begin{equation}
M=(0.291\pm 0.031)*(T[\mathrm{keV}])^{1.62\pm 0.08}*E(z)^{-1.04\pm0.07} 10^{14} M_{\sun},
\end{equation}
\begin{equation}
L_X\!=\!(0.079\pm 0.008)*(T[\mathrm{keV}])^{2.70\pm 0.24}*E(z)^{-0.23^{+0.12}_{-0.62}} 10^{44} \mathrm{erg\ s^{-1}},
\end{equation}
\begin{equation}
M\!=\!(1.39\pm 0.07)*(L_X[10^{44} \mathrm{erg\ s^{-1}}])^{0.54\pm 0.03}*E(z)^{-0.93^{+0.62}_{-0.12}} 10^{14} M_{\sun}.
\end{equation}

\noindent This work once again highlights the importance of selection biases to scaling relation studies. For distant cluster mass estimates based on the bolometric X-ray luminosity $L_X$, we recommend using Equ.\,\ref{eq:mlbiascorr}, which is based on the bias-corrected local relation of Pratt09 and includes a correction for the estimated selection bias on the observed evolution, given by   

\begin{equation}
M\!=\!(1.64\pm 0.07)*(L_X[10^{44} \mathrm{erg\ s^{-1}}])^{0.52\pm 0.03}*E(z)^{-0.90^{+0.35}_{-0.15}} 10^{14} M_{\sun}
\label{eq:mlbiascorr}
\end{equation}

\noindent To provide tighter constraints on scaling-relation evolution and improve the mass estimate of Equ.\,\ref{eq:mlbiascorr} in the future, a more homogeneous, extensive distant cluster sample with a precisely known selection function is necessary. Such a sample would allow a more precise bias estimate and correction to be made than possible for the sample used in this study.

Comparing the observed evolution with predictions made by the Millennium Gas Simulations has allowed us to discriminate between different proposed scenarios and attempt a physical interpretation of the thermodynamic history of the ICM. The comparison analysis strongly suggests an early preheating, \ie \ an entropy increase for the gas particles before the infall of the ICM gas into the cluster potential well. 

The expected number of cluster detections for the upcoming eROSITA survey was recalculated taking into account the results of this work. In general, the total number of achievable detections is slightly lower than assumed before, while the number of high redshift clusters to be detected shows a significant decrease.
  
Future more detailed studies of the redshift evolution of X-ray scaling relations will be important to more tightly constrain the early thermodynamic history of the ICM and provide calibrated mass-observable relations for upcoming large cluster surveys and their cosmological applications.

\bigskip

\begin{acknowledgements} 

We acknowledge Ben Maughan for helpful suggestions and the XDCP collaboration for providing some data prior to publication. We thank the anonymous referee for helpful comments and suggestions. The work of this paper was supported by the DfG cluster of excellence Origin and Structure of the Universe, by the German DLR under grant no. 50 QR 0802 and by the DfG under grant no. BO 702/17-3.

\end{acknowledgements}

\bibliographystyle{aa}
\bibliography{astrings,DA_BIB} 


\begin{appendix}

\section{Clusters included in the combined sample}
\label{sec:ClustersIncludedInTheCombinedSample}

\onecolumn
\topcaption{ Clusters included in the combined sample and their X-ray properties within $r_{500}$. T indicates the global ICM temperature, $L_X$ the bolometric X-ray luminosity and  M the (hydrostatic) cluster mass. Cluster names printed in boldface designate the additional $z\!>\!0.8$-clusters added to the sample (see Sect.\,\ref{sec:TheClusterSample}).} \label{tab:clusterlist}
\tablefirsthead{{\bf Cluster}   & {\bf $z$} & {\bf $kT[\mathrm{keV}]$} & {\bf $L_{\mathrm{X}}[10^{44} \mathrm{\ erg \ s^{-1}}]$} & {\bf $M[10^{14}M_{\sun}]$} & {\bf Source publication} \\
\hline }
\tablehead{
\multicolumn{6}{l}{\small\sl continued from previous page}\\
\hline
{\bf Cluster}   & {\bf $z$} & {\bf $kT[\mathrm{keV}]$} & {\bf $L_{\mathrm{X}}[10^{44} \mathrm{\ erg \ s^{-1}}]$} & {\bf $M[10^{14}M_{\sun}]$} & {\bf Source publication} \\
\hline }
\tabletail{%
\hline
\multicolumn{6}{r}{\small\sl continued on next page}\\}
\tablelasttail{\hline}
\begin{supertabular}{cccccc}
\hline

{\bf XMMXCSJ2215.9-1738}       & 1.46  & $4.1^{+0.6}_{-0.9}$          & $2.23^{+0.24}_{-0.35}$   &   & \citet{2010ApJ...718..133H}     \\
{\bf ISCSJ1438.1+3414}         & 1.41  & $3.3^{+1.9}_{-1.0}$          & $2.23^{+0.67}_{-0.53}$    &   & \citet{2010arXiv1012.0581B}     \\
{\bf XMMUJ2235.3-2557}         & 1.39  & $8.6^{+1.3}_{-1.2}$    & $10.0\pm 0.8$    & $4.4\pm 1$   & \citet{2009AA...508..583R}     \\
RX J0848+4453    & 1.270         & $2.9\pm 0.8$          & $1.0\pm 0.7$         & $1.2\pm 0.9$         & Ettori04        \\
RX J0849+4452    & 1.260         & $5.2\pm 1.6$          & $2.8\pm 0.2$         & $2.7\pm 1.4$         & Ettori04        \\
RX J1252--2927   & 1.235         & $5.2\pm 0.7$          & $5.8\pm 1.0$          & $1.5\pm 0.3$         & Ettori04        \\
{\bf XLSSJ022303.0-043622}     & 1.22  & $3.8^{+\infty}_{-1.9}$   & $1.1\pm 0.7$          &  &\citet{2006MNRAS.371.1427B}  \\
{\bf RXJ1053.7+5735}   & 1.134         & $3.9\pm 0.2$          & $2.2\pm 0.7$         &      & \citet{2004cosp...35.1588H}     \\
{\bf SPT-CLJ2106-5844}         & 1.132         & $8.5^{+2.6}_{-1.9}$          & $73.1\pm 5.3$    &      &  \citet{2011arXiv1101.1286F} \\
RX J0910+5422    & 1.106         & $6.6\pm 1.7$          & $2.8\pm 0.3$          & $4.6\pm 2.8$         & Ettori04        \\
{\bf XMMUJ100750.5+125818}     & 1.082         & $5.7^{+\infty}_{-3.65}$   & $1.3^{+\infty}_{-0.5}$   &      & \citet{Schwope2010a}    \\
SPTJ0546-5345    & 1.0665        & $7.5\pm 1.7$          & $26.9\pm 1.7$         &      & Andersson10     \\
{\bf 3C186}    & 1.067         & $5.58^{+0.28}_{-0.27}$    & $13.5\pm0.7$  & $2.31^{+0.21}_{-0.14}$   & \citet{2010ApJ...722..102S}    \\
{\bf XLSSJ022404.1-041330}     & 1.050  & $4.1\pm 0.9$  & $3.4^{+0.3}_{-0.2}$   & $1.3^{+0.9}_{-0.3}$   & \citet{2008MNRAS.387..998M}     \\
XLSSJ022709.2-041800     & 1.05  & $3.7\pm 1.5$          & $1.9\pm 0.2$   & $1.3\pm 0.3$   & Pacaud07        \\
ClJ1415.1+3612   & 1.03  & $5.7\pm 1.2$          & $10.1\pm 0.6$         & $3.0\pm 0.9$         & Maughan06       \\
SPTJ2341-5119    & 0.9983        & $8.0\pm 1.9$          & $24.5\pm 1.3$         &      & Andersson10     \\
{\bf 2XMMJ083026.2+524133}   & 0.99  & $8.2\pm 0.9$          & $18.5\pm 3.6$         & $5.6\pm 1$   & \citet{2008AA...487L..33L}     \\
{\bf XMMUJ1229.5+0151} & 0.975         & $6.4^{+0.7}_{-0.6}$          & $8.8\pm 1.5$&      &  \citet{2009AA...501...49S}   \\
{\bf XMMUJ1230.3+1339}    & 0.975         & $5.3^{+0.7}_{-0.6}$     & $6.5\pm0.7$    & $3.1\pm 1.8$     &  \citet{2011AA...527A..78F}    \\
ClJ1429.0+4241a  & 0.92  & $6.2\pm 1.5$          & $9.3\pm 0.9$    & $3.5\pm 1.3$  & Maughan06       \\
RCS2319+0030     & 0.904         & $6\pm 1$      & $7.9\pm 0.7$          & $2.8\pm 0.3$  & Hicks08 \\
{\bf CLJ1604+4304}  & 0.90  & $2.51^{+1.05}_{-0.69}$  & $2\pm 0.4$    &      & \citet{2004ApJ...601L...9L}     \\
RCS2320+0033     & 0.901         & $6.0\pm 2$    & $5.9\pm 0.5$          & $2.5\pm 0.5$        & Hicks08 \\
RCS2319+0038     & 0.900         & $5.9\pm 0.8$          & $16.2\pm 0.6$         & $3.6\pm 0.3$         & Hicks08 \\
WGA1226+3333     & 0.890         & $11.2\pm 2.2$         & $54.2\pm 0.8$         & $9.1\pm 2.4$         & Ettori04        \\
SPTJ0533-5005a   & 0.8810        & $4.0\pm 1.9$          & $3.6\pm 0.9$          &      & Andersson10     \\
ClJ1008.7+5342   & 0.87  & $3.6\pm 0.8$          & $3.6\pm 0.3$         & $1.5\pm 0.6$   & Maughan06       \\
RCS1620+2929     & 0.870         & $3.9\pm 1.2$          & $3.3\pm 0.5$          & $2.1\pm 0.4$        & Hicks08 \\
{\bf RXJ1257.2+4738}   & 0.866  & $3.6^{+2.9}_{-1.2}$          & $2.0^{+2.9}_{-1.2}$    &      & \citet{2009AA...503..399U}     \\
ClJ1559.1+6353a  & 0.85  & $4.1\pm 1.4$          & $2.5\pm 0.3$         & $1.6\pm 1.1$  & Maughan06       \\
XLSSJ022738.3-031758     & 0.84  & $3.3\pm 1.1$          & $4.1\pm 0.4$   & $0.9\pm 0.2$   & Pacaud07        \\
RX J0152--1357N  & 0.835         & $6.0\pm 1.1$          & $10.2\pm 0.6$         & $2.4\pm 0.7$        & Ettori04        \\
RX J0152--1357S  & 0.830         & $6.9\pm 2.9$          & $7.5\pm 0.4$          & $3.2\pm 1.4$         & Ettori04        \\
MS1054--0321     & 0.830         & $10.2\pm 1.0$         & $28.4\pm 3.0$   & $19.1\pm 3.5$         & Ettori04        \\
ClGJ1056-0337    & 0.826         & $9.2\pm 1.5$          & $49.1\pm 1.2$          &      & OHara07 \\
{\bf RXJ1821.6+6827}   & 0.816   & $4.7^{+1.2}_{-0.7}$   & $10.4^{+1.3}_{-1.7}$        &      & \citet{2004AA...428..867G}     \\
RX J1716+6708    & 0.813         & $6.8\pm 1.0$          & $13.6\pm 1.0$          & $4.0\pm 0.8$        & Ettori04        \\
RX J1350+6007    & 0.810         & $4.6\pm 0.7$          & $4.2\pm 0.4$         & $1.4\pm 0.4$         & Ettori04        \\
RX J1317+2911    & 0.805         & $4.1\pm 1.2$          & $1.1\pm 0.1$         & $1.4\pm 0.6$   & Ettori04        \\
XLSSJ022210.7-024048     & 0.79  & $3.9\pm 2.8$          & $1.4\pm 0.2$          & $1.8\pm 0.4$    & Pacaud07        \\
ClJ1103.6+3555a  & 0.78  & $6.0\pm 0.9$          & $4.5\pm 0.2$         & $2.9\pm 0.7$        & Maughan06       \\
RCS2318+0034     & 0.78  & $5.8\pm 1.2$          & $7.7\pm 0.8$      & $12.9\pm 2.0$    & Hicks08 \\
MS1137+6625      & 0.782         & $6.9\pm 0.5$          & $15.2\pm 0.4$         & $4.7\pm 0.6$        & Ettori04        \\
SPTJ2337-5942    & 0.7814        & $8.9\pm 2.0$          & $41.3\pm 2.3$        &      & Andersson10     \\
RCS0224-0002     & 0.778 & $5.1\pm 1.2$ & $4.1\pm 0.5$      & $5.0\pm 0.7$      & Hicks08 \\
XLSSJ022532.2-035511     & 0.77  & $2.8\pm 0.8$          & $2.1\pm 0.2$          & $0.9\pm 0.2$   & Pacaud07        \\
SPTJ0528-5300a   & 0.7648        & $5.2\pm 3.5$          & $6.6\pm 0.7$          &      & Andersson10     \\
RCS1107-0523     & 0.735         & $4.2\pm 0.6$          & $3.3\pm 0.3$      & $1.8\pm 0.2$     & Hicks08 \\
RX J2302+0844    & 0.734         & $6.6\pm 1.5$          & $5.3\pm 0.2$         & $3.2\pm 0.7$   & Ettori04        \\
RX J1113-2615   & 0.730         & $5.6\pm 0.8$          & $4.4\pm 0.8$          & $3.2\pm 0.7$   & Ettori04        \\
ClJ1113.1-2615   & 0.72  & $4.7\pm 0.9$          & $3.7\pm 0.3$         & $2.6\pm 0.8$        & Maughan06       \\
ClJ1342.9+2828a  & 0.71  & $3.7\pm 0.5$          & $3.3\pm 0.1$         & $1.8\pm 0.4$         & Maughan06       \\
WJ 1342.8+4028   & 0.70  & $3.5\pm 0.3$          & $3.8\pm 0.8$         & $1.3\pm 0.2$         & Kotov05 \\
RX J1221+4918    & 0.700         & $7.5\pm 0.7$          & $12.7\pm 0.4$         & $5.5\pm 0.9$  & Ettori04        \\
ClGJ0744+3927    & 0.686         & $9.6\pm 0.9$          & $46.7\pm 0.7$          &      & OHara07 \\
ClGJ1419+5326    & 0.640         & $4.1\pm 0.8$          & $6.9\pm 0.2$   &      & OHara07 \\
RX J0542-4100   & 0.634         & $7.9\pm 1.0$          & $11.6\pm 1.3$          & $3.9\pm 0.8$         & Ettori04        \\
RCS1419+5326     & 0.62  & $4.6\pm 0.4$          & $8.4\pm 0.5$          & $2.6\pm 0.147$       & Hicks08 \\
RX J1334.3+5030  & 0.62  & $4.6\pm 0.4$          & $7.7\pm 1.5$          & $2.8\pm 0.5$         & Kotov05 \\
ClJ0046.3+8530   & 0.62  & $4.4\pm 0.5$          & $3.8\pm 0.2$         & $2.1\pm 0.4$         & Maughan06       \\
XLSSJ022457.1-034856     & 0.61  & $3.2\pm 0.4$          & $3.6\pm 0.2$   & $1.2\pm 0.3$   & Pacaud07        \\
SPTJ0559-5249    & 0.6112        & $7.7\pm 1.1$          & $14.1\pm 0.9$         &      & Andersson10     \\
RX J1120.1+4318  & 0.60  & $4.9\pm 0.3$          & $13.4\pm 2.7$          & $4.7\pm 1.2$         & Kotov05 \\
ClGJ0647+7015   & 0.584         & $15.0\pm 3.8$         & $49.9\pm 1.1$          &      & OHara07 \\
MS2053-0449    & 0.583         & $5.5\pm 0.5$          & $5.3\pm 1.0$          & $3.1\pm 0.5$        & Ettori04        \\
SPTJ2331-5051   & 0.5707        & $5.9\pm 1.3$          & $16.2\pm 0.7$         &      & Andersson10     \\
RX J0848+4456   & 0.570         & $3.2\pm 0.3$          & $1.2\pm 0.4$         & $1.4\pm 0.3$  & Ettori04        \\
ClGJ2129-0741   & 0.570         & $11.8\pm 2.8$         & $39.4\pm 1.3$          &      & OHara07 \\
RX J1121+2326   & 0.562         & $4.6\pm 0.5$          & $5.4\pm 0.2$          & $5.1\pm 1.8$         & Ettori04        \\
ClGJ0717+3745   & 0.548         & $11.5\pm 0.7$         & $121.8\pm 2.2$          &      & OHara07 \\
ClGJ1354-0221   & 0.546         & $4.1\pm 0.8$          & $5.9\pm 0.3$          &      & OHara07 \\
ClGJ1423+2404   & 0.545         & $5.4\pm 0.2$          & $37.2\pm 0.2$   &      & OHara07 \\
ClGJ1149+2223   & 0.544         & $9.8\pm 0.8$          & $64.1\pm 1.5$   &      & OHara07 \\
MS0016+1609     & 0.541         & $10.0\pm 0.5$         & $52.0\pm 7.2$          & $8.8\pm 0.7$   & Ettori04        \\
MS0451-0305    & 0.539         & $8.0\pm 0.3$          & $50.2\pm 7.9$          & $7.1\pm 0.5$        & Ettori04        \\
RX J1525+0957   & 0.516         & $5.1\pm 0.5$          & $6.6\pm 0.2$         & $2.8\pm 0.5$        & Ettori04        \\
RX J0505.3+2849         & 0.51  & $2.5\pm 0.4$          & $1.6\pm 0.3$         & $1.4\pm 0.2$         & Kotov05 \\
XLSSJ022206.7-030314    & 0.49  & $3.6\pm 0.6$          & $3.1\pm 0.2$   & $1.6\pm 0.3$    & Pacaud07        \\
XLSSJ022357.4-043517    & 0.49  & $2.2\pm 0.9$          & $0.46\pm 0.04$          & $0.5\pm 0.1$   & Pacaud07        \\
CL 1641+4001   & 0.464         & $5.1\pm 0.8$          & $3.0\pm 0.2$   &      & Branchesi07     \\
SPTJ0509-5342   & 0.4626        & $7.0\pm 1.4$          & $13.0\pm 0.4$   &      & Andersson10     \\
ClGJ1621+3810   & 0.461         & $6.8\pm 0.6$          & $20.2\pm 0.2$          &      & OHara07 \\
3C 295  & 0.460         & $4.3\pm 0.3$          & $13.7\pm 0.2$         & $2.0\pm 0.2$        & Ettori04        \\
RX J1701+6421   & 0.453         & $4.5\pm 0.4$          & $5.8\pm 0.5$    & $1.3\pm 0.1$         & Ettori04        \\
RX J1347-1145  & 0.451         & $10.3\pm 0.6$         & $114.8\pm 0.6$         & $7.8\pm 0.7$         & Ettori04        \\
ClGJ0329-0212   & 0.450         & $5.9\pm 0.2$          & $30.5\pm 0.2$          &      & OHara07 \\
MACSJ0417.5-1154        & 0.440         & $9.4\pm 0.7$          & $128.2\pm 2.0$          &      & OHara07 \\
RXCJ1206.2-0848         & 0.440         & $11.4\pm 0.9$         & $53.0\pm 0.6$          &      & OHara07 \\
XLSSJ022145.2-034617    & 0.43  & $4.8\pm 0.6$          & $6.6\pm 0.2$          & $2.8\pm 0.6$    & Pacaud07        \\
MS1621+2640     & 0.426         & $6.8\pm 0.9$          & $10.3\pm 0.3$          & $4.0\pm 0.8$        & Ettori04        \\
MS 0302.5+1717  & 0.42  & $4.5\pm 0.5$          & $4.4\pm 0.9$         & $2.2\pm 0.6$  & Kotov05 \\
MS0302+1658     & 0.424         & $3.8\pm 0.9$          & $6.3\pm 0.2$         & $2.1\pm 0.5$        & Ettori04        \\
SPTJ0551-5709b  & 0.4230        & $4.1\pm 0.9$          & $5.8\pm 0.6$          &      & Andersson10     \\
RXCJ2228.6+2036         & 0.412         & $8.1\pm 0.5$          & $35.2\pm 0.7$          &      & OHara07 \\
RX J1416+4446   & 0.400         & $3.7\pm 0.2$          & $5.0\pm 0.2$         & $1.2\pm 0.1$        & Ettori04        \\
Zw Cl 0024.0+1652     & 0.394         & $4.4\pm 0.5$          & $4.2\pm 0.2$   &      & Branchesi07     \\
RXCJ0949.8+1707         & 0.383         & $7.8\pm 0.7$          & $31.3\pm 0.5$          &      & OHara07 \\
Zw Cl 1953        & 0.374         & $7.6\pm 0.5$          & $26.2\pm 0.4$          &      & OHara07 \\
Abell 370    & 0.373         & $8.7\pm 0.5$          & $22.4\pm 0.4$          &      & OHara07 \\
RXJ1532.9+3021  & 0.362         & $6.1\pm 0.3$          & $43.8\pm 0.2$    &      & OHara07 \\
RXCJ0404.6+1109         & 0.355         & $5.6\pm 0.8$          & $9.8\pm 0.6$         &      & OHara07 \\
XLSSJ022722.4-032144    & 0.33  & $2.4\pm 0.5$          & $0.7\pm 0.1$   & $0.9\pm 0.2$   & Pacaud07        \\
Abell 1722   & 0.328         & $9.1\pm 1.5$          & $13.7\pm 0.3$          &      & OHara07 \\
Zw Cl 1358+6245   & 0.327         & $9.1\pm 0.9$          & $19.4\pm 0.3$          &      & OHara07 \\
XLSSJ022402.0-050525    & 0.32  & $2.0\pm 0.7$          & $0.14\pm 0.02$          & $0.7\pm 0.1$   & Pacaud07        \\
Abell 1995   & 0.318         & $8.1\pm 1.0$          & $17.1\pm 0.2$          &      & OHara07 \\
MS2137.3-2353   & 0.313         & $5.0\pm 0.2$          & $32.1\pm 0.2$          &      & OHara07 \\
Abell 1300        & 0.3075        & $9.2\pm 0.4$          & $18.0\pm 1.5$        & $5.2\pm 3.0$        & Zhang07 \\
Abell 2744        & 0.3066        & $10.1\pm 0.3$         & $22.12\pm 1.7$       & $7.4\pm 2.9$        & Zhang07 \\
RXCJ2245.0+2637         & 0.304         & $5.9\pm 0.3$          & $18.3\pm 0.1$          &      & OHara07 \\
MS 1008.1-1224         & 0.302         & $6.0\pm 0.4$          & $11.3\pm 0.4$         &      & Branchesi07     \\
Abell 781         & 0.298         & $6.5\pm 0.5$          & $6.3\pm 1$    & $4.5\pm 1.3$       & Zhang08 \\
RXC J2308.3-0211         & 0.297         & $7.6\pm 0.7$          & $12.1\pm 1.3$         & $7.4\pm 2.3$       & Zhang08 \\
Abell 2537               & 0.2966        & $7.6\pm 0.9$        & $15.8\pm 1.0$       & $7.2\pm 1.1$         & Mantz09 \\
1ES 0657-558             & 0.2965        & $11.7\pm 0.2$       & $65.2\pm 0.9$         & $22.8\pm 2.8$        & Mantz09 \\
RXC J0658.5-5556         & 0.296         & $10.7\pm 0.4$         & $49.5\pm 2.4$         & $11.0\pm 5.3$      & Zhang08 \\
RXC J0516.7-5430         & 0.294         & $6.7\pm 0.5$          & $10.9\pm 1.5$         & $6.4\pm 1.8$       & Zhang08 \\
XLSSJ022803.4-045103     & 0.29  & $2.8\pm 0.6$          & $0.50\pm 0.04$        & $1.4\pm 0.1$         & Pacaud07        \\
RXC J0043.4-2037         & 0.292         & $7.0\pm 0.4$          & $10.5\pm 1$   & $4.8\pm 1.4$       & Zhang08 \\
 Zw Cl 3146              & 0.2906        & $8.4\pm 0.4$        & $49.1\pm 1.8$        & $9.4\pm 1.2$         & Mantz09 \\
Abell 611     & 0.288         & $8.9\pm 0.7$          & $18.9\pm 0.3$          &      & OHara07 \\
RXC J0528.9-3927         & 0.284         & $6.6\pm 0.5$          & $14.5\pm 1.5$         & $6.4\pm 1.9$       & Zhang08 \\
RXC J0232.2-4420         & 0.284         & $6.6\pm 0.3$          & $18.5\pm 1.4$         & $8.4\pm 2.5$       & Zhang08 \\
RXCJ0437.1+0043  & 0.2842        & $5.1\pm 0.3$          & $6.2\pm 0.7$         & $6.1\pm 2.2$        & Zhang07 \\
Abell 697                & 0.282         & $10.9\pm 1.1$       & $41.9\pm 2.3$        & $17.1\pm 2.9$        & Mantz09 \\
Abell 1758        & 0.280         & $7.9\pm 0.2$          & $10.5\pm 0.7$         & $11.2\pm 3.4$      & Zhang08 \\
RX J2011.3-5725          & 0.2786        & $3.2\pm 0.3$        & $6.0\pm 0.3$   & $3.3\pm 0.7$         & Mantz09 \\
RXC J0532.9-3701         & 0.275         & $7.7\pm 0.6$          & $12.7\pm 1.2$         & $5.4\pm 1.6$       & Zhang08 \\
RXC J2337.6+0016         & 0.275         & $7.5\pm 0.5$          & $10.0\pm 0.9$   & $11.0\pm 3.2$      & Zhang08 \\
RXCJ0303.7-7752  & 0.2742        & $8.2\pm 0.5$          & $12.9\pm 1.3$        & $7.7\pm 2.3$        & Zhang07 \\
XLSSJ022524.7-044039     & 0.26  & $2.0\pm 0.2$          & $0.49\pm 0.02$       & $0.6\pm 0.1$         & Pacaud07        \\
MS 1006.01202     & 0.261         & $6.1\pm 0.4$          & $10.8\pm 0.2$         &      & OHara07 \\
RXC J0307.0-2840         & 0.258         & $7.1\pm 0.4$          & $13.1\pm 1.2$         & $5.5\pm 2.0$       & Zhang08 \\
Zw Cl 7160    & 0.258         & $4.7\pm 0.1$          & $17.0\pm 0.8$   & $2.4\pm 0.7$       & Zhang08 \\
MS J1455.0+2232          & 0.2578        & $4.5\pm 0.2$        & $20.7\pm 0.6$       & $6.2\pm 1.0$         & Mantz09 \\
Abell 68  & 0.255         & $7.3\pm 0.3$          & $11.4\pm 1$   & $6.5\pm 1.9$       & Zhang08 \\
Abell 1835        & 0.253         & $8.4\pm 0.3$          & $53.2\pm 1.7$         & $8.0\pm 2.3$       & Zhang08 \\
Abell 521                & 0.2475        & $6.2\pm 0.3$        & $15.7\pm 0.7$        & $11.4\pm 1.7$        & Mantz09 \\
Abell 2125      & 0.246         & $3.4\pm 0.2$          & $1.9\pm 0.1$        &      & Branchesi07     \\
RX J0439.0+0715          & 0.2443        & $6.6\pm 0.5$        & $18.4\pm 0.7$       & $7.4\pm 1.0$         & Mantz09 \\
RXC J2129.6+0005         & 0.235         & $6.3\pm 0.2$          & $14.3\pm 0.9$         & $4.3\pm 1.3$       & Zhang08 \\
 Zw Cl 2089              & 0.2347        & $6.6\pm 1.5$        & $13.2\pm 0.7$       & $3.1\pm 0.4$         & Mantz09 \\
Abell 2390        & 0.233         & $11.6\pm 0.6$         & $40.9\pm 2.7$         & $7.7\pm 2.3$       & Zhang08 \\
Abell 2667        & 0.230         & $7.0\pm 0.3$          & $21.6\pm 1.3$         & $6.0\pm 1.7$       & Zhang08 \\
Abell 267         & 0.230         & $6.2\pm 0.4$          & $7.7\pm 0.7$          & $4.3\pm 1.3$       & Zhang08 \\
Abell 2111               & 0.229         & $6.5\pm 0.7$        & $11.3\pm 0.7$       & $7.8\pm 1.9$         & Mantz09 \\
 Zw Cl 5247              & 0.229         & $5.3\pm 1.1$        & $8.8\pm 0.6$        & $8.2\pm 1.8$         & Mantz09 \\
Abell1763        & 0.228         & $5.8\pm 0.3$          & $16.5\pm 1.5$         & $5.0\pm 1.5$       & Zhang08 \\
RX J0220.9-3829          & 0.228         & $5.2\pm 0.5$        & $8.0\pm 0.4$         & $4.4\pm 1.1$         & Mantz09 \\
Abell 2219               & 0.2281        & $10.9\pm 0.5$       & $45.1\pm 2.3$       & $18.9\pm 2.5$        & Mantz09 \\
Abell 1682               & 0.226         & $7.0\pm 2.1$        & $15.3\pm 1.7$       & $12.4\pm 3.2$        & Mantz09 \\
RX J0638.7-5358          & 0.2266        & $9.5\pm 1.0$        & $30.6\pm 1.6$       & $10.3\pm 1.4$        & Mantz09 \\
Abell 2261        & 0.224         & $6.6\pm 0.6$          & $13.9\pm 2.2$         & $6.0\pm 1.7$       & Zhang08 \\
RX J0237.4-2630          & 0.2216        & $6.7\pm 1.3$        & $12.8\pm 0.7$       & $5.6\pm 1.1$         & Mantz09 \\
RX J0304.1-3656          & 0.2192        & $6.3\pm 0.8$        & $6.7\pm 0.4$        & $4.4\pm 0.9$         & Mantz09 \\
Abell 773         & 0.217         & $8.3\pm 0.4$          & $20.9\pm 1.6$         & $8.3\pm 2.5$       & Zhang08 \\
RX J1504.1-0248          & 0.2153        & $8.0\pm 0.4$        & $69.4\pm 2.5$        & $11.0\pm 1.4$        & Mantz09 \\
 Zw Cl 2701              & 0.214         & $6.8\pm 0.5$        & $10.4\pm 0.5$       & $4.0\pm 0.7$         & Mantz09 \\
Abell 1423               & 0.213         & $5.8\pm 0.6$        & $13.2\pm 0.9$       & $8.7\pm 2.0$         & Mantz09 \\
Abell 209         & 0.209         & $7.1\pm 0.3$          & $13.3\pm 1.1$         & $5.3\pm 1.7$       & Zhang08 \\
RX J0439.0+0520          & 0.208         & $5.0\pm 0.5$        & $8.7\pm 0.2$         & $2.7\pm 0.5$         & Mantz09 \\
Abell 963         & 0.206         & $6.5\pm 0.2$          & $11.4\pm 0.9$         & $5.2\pm 1.5$       & Zhang08 \\
Abell 520                & 0.203         & $7.2\pm 0.2$        & $20.1\pm 0.7$       & $11.9\pm 1.6$        & Mantz09 \\
Abell 2163               & 0.203         & $12.3\pm 0.9$       & $88.1\pm 3.4$        & $38.5\pm 5.0$        & Mantz09 \\
Abell 115         & 0.197         & $6.2\pm 0.1$          & $14.3\pm 1.1$         & $4.2\pm 1.1$       & Zhang08 \\
Abell 383         & 0.187         & $4.7\pm 0.2$          & $8.1\pm 0.5$          & $3.2\pm 0.9$       & Zhang08 \\
Abell 1689        & 0.184         & $8.5\pm 0.2$          & $28.4\pm 1$   & $10.3\pm 3.0$      & Zhang08 \\
RXC J1311.4-0120         & 0.1832        & $8.9\pm 0.1$        & $36.1\pm 0.1$      &      & Pratt09 \\
Abell 665                & 0.1818        & $8.0\pm 0.2$        & $21.7\pm 2.0$       & $12.7\pm 1.8$        & Mantz09 \\
Abell 2218        & 0.176         & $6.6\pm 0.3$          & $11.1\pm 0.8$         & $4.2\pm 1.3$       & Zhang08 \\
Abell 1914        & 0.171         & $8.8\pm 0.3$          & $21.7\pm 1.1$         & $16.8\pm 4.9$      & Zhang08 \\
RXC J0958.3-1103         & 0.167         & $5.8\pm 0.3$          & $9.1\pm 0.9$          & $3.3\pm 1.0$       & Zhang08 \\
RXC J0645.4-5413         & 0.164         & $7.6\pm 0.3$          & $17.8\pm 1.6$         & $6.6\pm 2.0$       & Zhang08 \\
Abell 901         & 0.163         & $3.2\pm 0.2$          & $16\pm 0.5$   & $3.2\pm 1.0$       & Zhang08 \\
Abell 907     & 0.1603        & $5.96\pm 0.08$        &       & $4.7\pm 0.4$         & Vikhlinin06     \\
RXC J2014.8-2430         & 0.1538        & $4.8\pm 0.1$        & $21.1\pm 0.1$      &      & Pratt09 \\
RXC J0945.4-0839         & 0.153         & $5.3\pm 0.5$          & $4.2\pm 0.6$          & $6.7\pm 2.0$       & Zhang08 \\
Abell 2204        & 0.152         & $7.6\pm 0.2$          & $33.9\pm 1.2$         & $5.3\pm 1.5$       & Zhang08 \\
RXC J2234.5-3744         & 0.151         & $7.5\pm 0.4$          & $10.9\pm 1.1$         & $8.1\pm 2.4$       & Zhang08 \\
RXC J2217.7-3543         & 0.1486        & $4.9\pm 0.1$        & $6.12\pm 0.03$       &      & Pratt09 \\
RXC J0547.6-3152         & 0.148         & $6.0\pm 0.2$          & $7.2\pm 0.6$          & $5.8\pm 1.7$       & Zhang08 \\
RXC J2048.1-1750         & 0.1475        & $4.7\pm 0.1$        & $5.13\pm 0.03$       &      & Pratt09 \\
Abell 1413        & 0.143         & $6.6\pm 0.1$          & $13.8\pm 0.8$         & $5.4\pm 1.6$       & Zhang08 \\
XLSSJ022540.6-031121     & 0.14  & $3.5\pm 0.6$          & $1.0\pm 0.1$        & $2.4\pm 0.1$  & Pacaud07        \\
RXC J2218.6-3853         & 0.141         & $6.2\pm 0.2$          & $6.6\pm 0.5$          & $4.8\pm 1.4$       & Zhang08 \\
RXC J0020.7-2542         & 0.1410        & $5.7\pm 0.1$        & $6.52\pm 0.04$       &      & Pratt09 \\
RXC J0605.8-3518         & 0.1392        & $4.6\pm 0.1$        & $9.54\pm 0.04$       &      & Pratt09 \\
Abell 1068               & 0.1386        & $3.9\pm 0.1$        & $9.7\pm 0.1$        & $3.7\pm 0.6$         & Mantz09 \\
RXC J1044.5-0704         & 0.1342        & $3.41\pm 0.03$        & $7.42\pm 0.02$       &      & Pratt09 \\
MS J1111.8               & 0.1306        & $5.8\pm 0.2$        & $6.8\pm 0.6$        & $4.5\pm 0.7$         & Mantz09 \\
RXC J1516.5-0056         & 0.1198        & $3.6\pm 0.1$        & $2.31\pm 0.02$       &      & Pratt09 \\
RXC J1141.4-1216         & 0.1195        & $3.31\pm 0.03$        & $3.75\pm 0.01$       &      & Pratt09 \\
RXC J2149.1-3041         & 0.1184        & $3.26\pm 0.04$        & $3.56\pm 0.02$       &      & Pratt09 \\
RXC J1516.3+0005         & 0.1181        & $4.5\pm 0.1$        & $4.12\pm 0.02$       &      & Pratt09 \\
RXC J0145.0-5300         & 0.1168        & $5.5\pm 0.1$        & $5.00\pm 0.03$       &      & Pratt09 \\
RXC J0616.8-4748         & 0.1164        & $4.2\pm 0.1$        & $2.38\pm 0.02$       &      & Pratt09 \\
RXC J0006.0-3443         & 0.1147        & $5.0\pm 0.2$        & $4.1\pm 0.1$       &      & Pratt09 \\
Abell 2034               & 0.113         & $7.2\pm 0.3$        & $9.5\pm 1.0$        & $6.7\pm 1.0$         & Mantz09 \\
RXC J0049.4-2931         & 0.1084        & $3.1\pm 0.1$        & $1.78\pm 0.02$       &      & Pratt09 \\
PKS 0745-191     & 0.1028        & $8.0\pm 0.3$        &       & $7.3\pm 0.8$       & Arnaud05        \\
RXC J0211.4-4017         & 0.1008        & $2.1\pm 0.1$        & $0.81\pm 0.01$       &      & Pratt09 \\
Abell 2244               & 0.0989        & $5.4\pm 0.1$        & $10.7\pm 1.0$       & $6.2\pm 1.1$         & Mantz09 \\
RXC J2319.6-7313         & 0.0984        & $2.22\pm 0.03$        & $2.00\pm 0.02$       &      & Pratt09 \\
Abell 3921               & 0.094         & $5.1\pm 0.2$        & $6.2\pm 0.6$         & $5.4\pm 0.9$         & Mantz09 \\
RXC J0003.8+0203         & 0.0924        & $3.9\pm 0.1$        & $1.88\pm 0.01$       &      & Pratt09 \\
Abell 2142               & 0.0904        & $10.0\pm 0.3$       & $34.8\pm 1.7$       & $13.9\pm 2.1$        & Mantz09 \\
Abell 478                & 0.0881        & $8.0\pm 0.3$        & $33.4\pm 3.0$       & $10.1\pm 1.6$        & Mantz09 \\
Abell 2597               & 0.0852        & $3.6\pm 0.1$        & $7.2\pm 0.7$        & $2.9\pm 0.5$         & Mantz09 \\
RXC J1302.8-0230         & 0.0847        & $3.0\pm 0.1$        & $1.38\pm 0.01$       &      & Pratt09 \\
RXC J0821.8+0112         & 0.0822        & $2.7\pm 0.1$        & $0.77\pm 0.01$       &      & Pratt09 \\
Abell 2255               & 0.0809        & $6.4\pm 0.2$        & $6.5\pm 0.7$        & $5.9\pm 1.0$         & Mantz09 \\
RXC J2129.8-5048         & 0.0796        & $3.8\pm 0.2$        & $1.46\pm 0.02$       &      & Pratt09 \\
RXC J1236.7-3354         & 0.0796        & $2.7\pm 0.1$        & $1.03\pm 0.01$       &      & Pratt09 \\
Abell 2029               & 0.0779        & $8.2\pm 0.2$        & $27.0\pm 2.6$       & $9.3\pm 1.4$         & Mantz09 \\
Abell 3112               & 0.0752        & $4.3\pm 0.1$        & $8.0\pm 0.7$   & $4.1\pm 0.6$         & Mantz09 \\
Abell 401                & 0.0743        & $7.7\pm 0.3$        & $16.8\pm 1.0$       & $10.1\pm 1.6$        & Mantz09 \\
Abell 1795               & 0.0622        & $6.1\pm 0.1$        & $13.2\pm 0.9$        & $5.5\pm 0.8$         & Mantz09 \\
RXC J0225.1-2928         & 0.0604        & $2.5\pm 0.2$        & $0.51\pm 0.01$       &      & Pratt09 \\
RXC J0345.7-4112         & 0.0603        & $2.19\pm 0.04$        & $0.77\pm 0.01$       &      & Pratt09 \\
Abell 3266               & 0.0602        & $8.6\pm 0.2$        & $12.8\pm 0.8$       & $9.2\pm 1.4$         & Mantz09 \\
Abell 1991   & 0.0586        & $2.7\pm 0.1$        &       & $1.2\pm 0.1$       & Arnaud05        \\
Abell 2256               & 0.0581        & $6.9\pm 0.2$        & $10.7\pm 0.9$       & $7.2\pm 1.0$         & Mantz09 \\
RXC J2157.4-0747         & 0.0579        & $2.5\pm 0.1$        & $0.45\pm 0.01$       &      & Pratt09 \\
Abell 133     & 0.0569        & $4.1\pm 0.1$        &       & $3.3\pm 0.4$         & Vikhlinin06     \\
RXC J2023.0-2056         & 0.0564        & $2.7\pm 0.1$        & $0.61\pm 0.01$       &      & Pratt09 \\
Abell 85                 & 0.0557        & $6.5\pm 0.1$        & $12.8\pm 0.9$        & $7.2\pm 1.0$         & Mantz09 \\
Abell 3667               & 0.0557        & $6.3\pm 0.1$        & $13.0\pm 1.1$       & $11.8\pm 2.2$        & Mantz09 \\
Abell 2717   & 0.0498        & $2.6\pm 0.1$        &       & $1.1\pm 0.1$       & Arnaud05        \\
Abell 3558               & 0.048         & $5.5\pm 0.1$        & $7.7\pm 0.6$        & $6.4\pm 1.0$         & Mantz09 \\
Abell 1983   & 0.0442        & $2.2\pm 0.1$        &       & $1.1\pm 0.4$       & Arnaud05        \\
MKW 9    & 0.0382        & $2.4\pm 0.2$        &       & $0.9\pm 0.2$       & Arnaud05        \\
\hline
\end{supertabular}
\twocolumn

\section{L-z distribution of the cluster sample}
\label{sec:LXZDistributionOfTheClusterSample}

\begin{figure}[ht]
	\flushleft
		\includegraphics[angle=-90,width=0.5\textwidth]{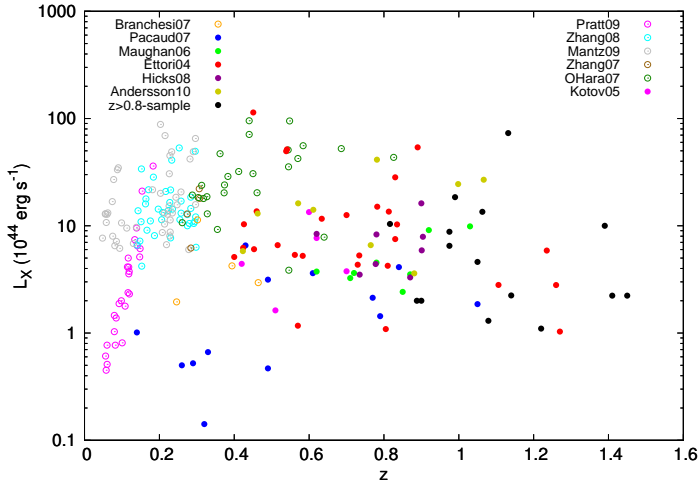}
	\caption{Bolometric X-ray luminosity $L_X$ of the clusters included in the combined cluster sample.}
	\label{fig:lz}
\end{figure}

\section{Comparison of cluster properties for systems included in more than one subsample}
\label{sec:ComparisonOfClusterPropertiesForSystemsIncludedInMoreThanOneSubsample}

\begin{figure}[ht]
	\centering
		\includegraphics[angle=-90,width=0.5\textwidth]{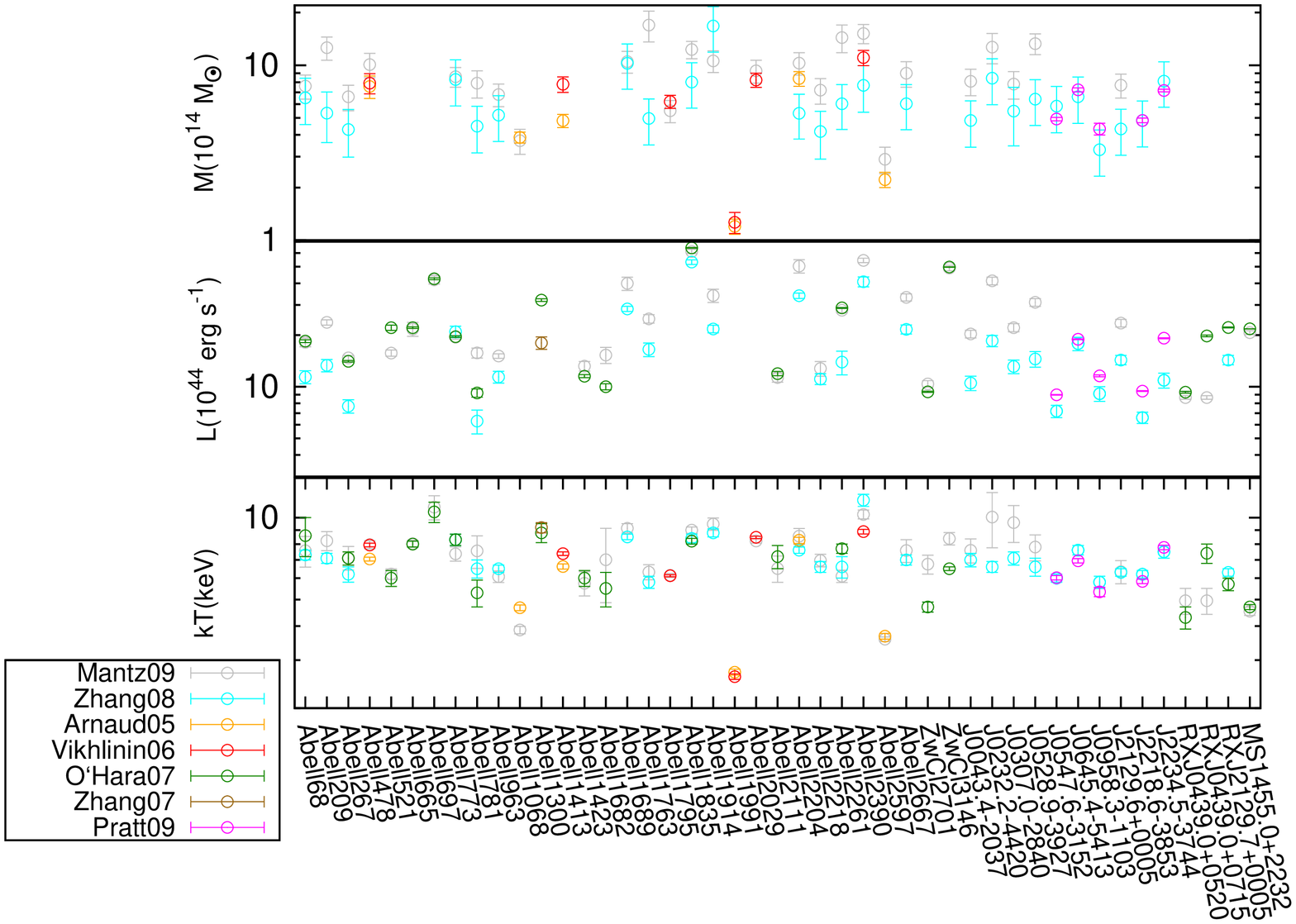}
	\caption{Comparison of cluster properties for $z\!<\! 0.3$-systems included in more than one subsample.}
	\label{fig:sharedlowz}
\end{figure}

\begin{figure}[ht]
	\centering
		\includegraphics[angle=-90,width=0.5\textwidth]{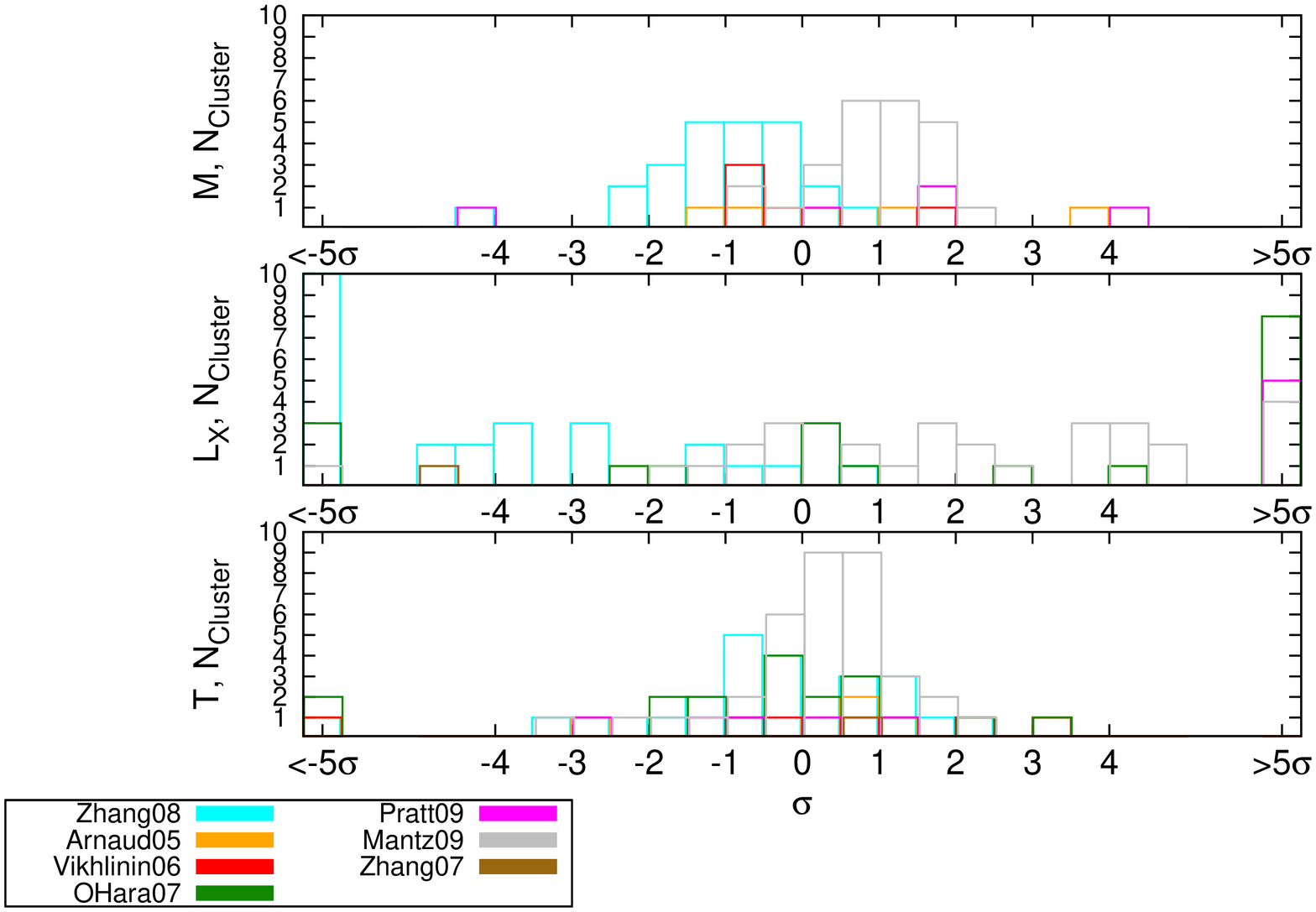}
	\caption{Distribution of deviations from the mean value for the $z\!<\! 0.3$-clusters included in more than one subsample in units of the assumed error $\sigma$.  Top panel: Mass deviations. Middle panel: $L_X$ deviations. Bottom panel: ICM temperature deviations.}
	\label{siglowz}
\end{figure}

\begin{figure}[ht]
	\centering
		\includegraphics[angle=-90,width=0.5\textwidth]{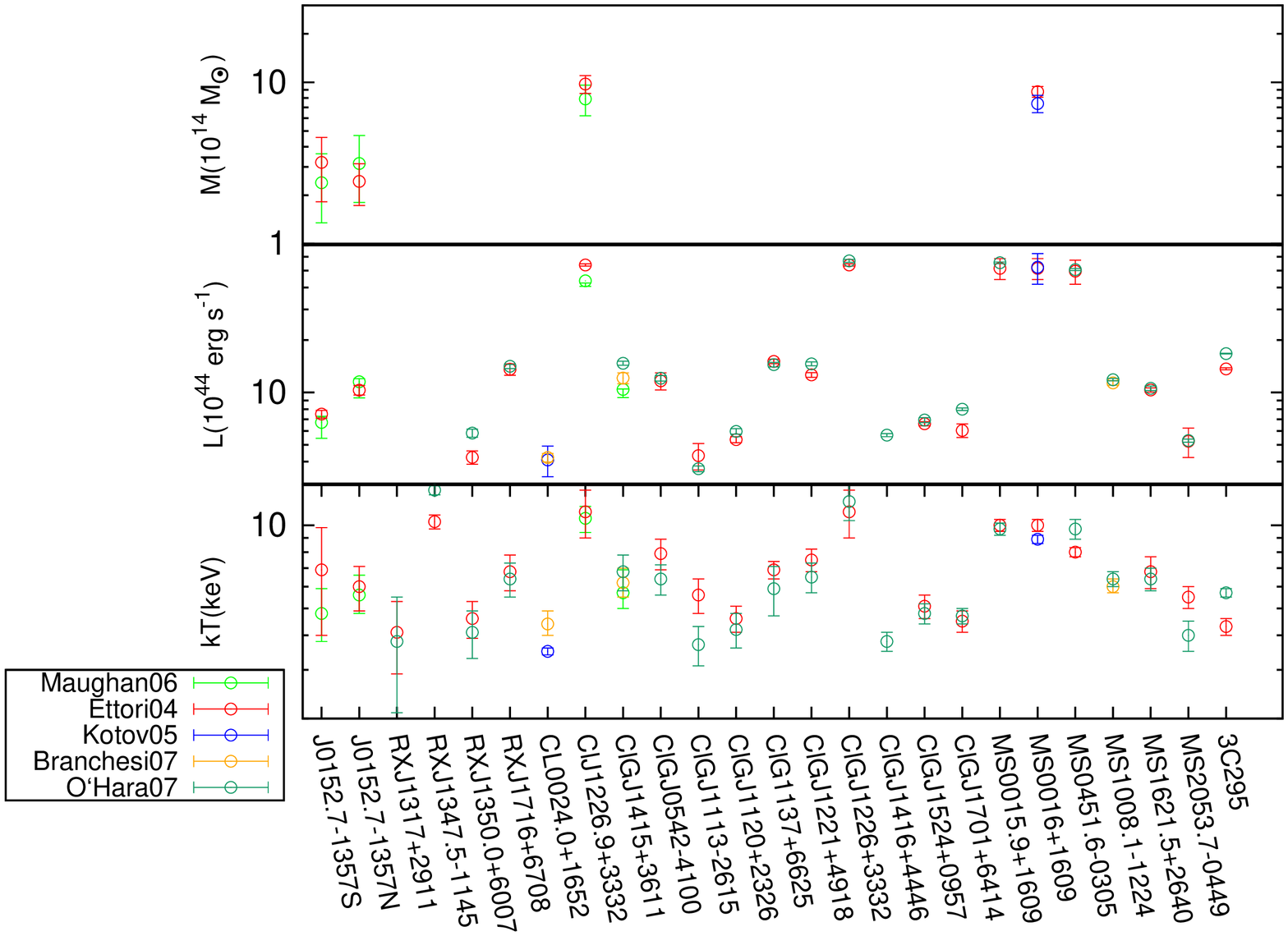}
	\caption{Comparison of cluster properties for $z\!>\! 0.3$-systems included in more than one subsample.}
	\label{fig:sharedhighz}
\end{figure}

\begin{figure}[ht]
	\centering
		\includegraphics[angle=-90,width=0.5\textwidth]{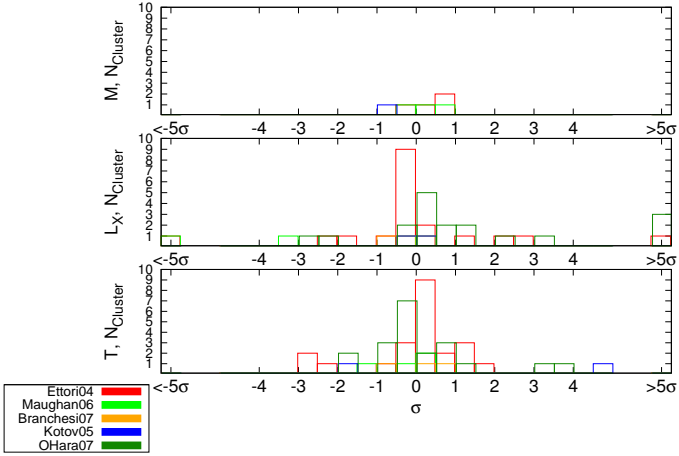}
	\caption{Distribution of deviations from the mean value for the $z\!>\! 0.3$-clusters included in more than one subsample in units of the assumed error $\sigma$.  Top panel: Mass deviations. Middle panel: $L_X$ deviations. Bottom panel: ICM temperature deviations.}
	\label{sighighz}
\end{figure}

Several of the clusters in our combined sample are included in more than one source publication, either relying on completely independent measurement data or using the same X-ray data reanalyzed according to the strategy that was chosen by the different authors. These measurements of cluster observables can be compared to each other after correcting for different analysis schemes and provide a useful tool to test the applied homogenization scheme. Furthermore, the comparison analysis gives an estimate of whether the error budget assumed in the different publications is realistic and reveals systematic differences between the results derived by the various studies.

Fig.\,\ref{fig:sharedlowz} shows the cluster properties of the low redshift ($z\!<\!0.3$) overlap sample. In Fig.\,\ref{siglowz}, we present the deviations of the individual measured values from the mean value in units of $\sigma$, and the error estimated in the individual publications. Note that the true values of the cluster observables are unknown. The comparison to the mean value therefore does not allow any statements about the reliability of the results derived in the different studies.  Figure\,\ref{siglowz} instead provides an insight into the systematic differences between the results of different studies and whether the assumed error estimates are realistic.

The derived spectroscopic temperatures agree well for most of the clusters. As visible in the bottom panel of Fig.\,\ref{siglowz}, the majority of the measured values deviate less than $1\sigma$ from the mean value. In detail, $59\%$ of the measurements lie within $1\sigma$ and $82\%$ within $2\sigma$ of the mean. Only $5\%$ of the results deviate by more than $5\sigma$. This indicates that the spectral fitting method generally leads to secure and consistent results, that the probability of severe misestimations is low, and that the assumed error budgets are likely to be realistic. Differences may result from different spectral extraction regions or different treatments of parameters, such as the ICM metallicity or the background subtraction process. However, these different measurement schemes do not lead to completely incomparable data sets. The temperature differences between the subsamples are rather uncorrelated and reveal no systematic trends between different studies. We note that for the samples of Zhang07, Zhang08, and Arnaud05 only core-excised temperatures were available. However, comparing those to the core-included temperatures given in other studies (\eg \ Mantz09), the observed differences remain small.
 
For the X-ray luminosity $L_X$, the situation is clearly different. As visible in the middle panel of Fig.\,\ref{fig:sharedlowz}, most of the derived luminosities do not agree within the errors. Furthermore, the differences between the results of some studies clearly show systematic trends. In terms of the deviations from the mean value, only $19\%$ of the values lie within $1\sigma$ and $29\%$ within $2\sigma$, while $39\%$ of the measurements show deviations of more than $5\sigma$. The different samples exhibit systematic differences when compared to each other, especially for the Mantz09-Zhang08 overlap but to a lesser degree also for the common clusters of Zhang08 and OHara07. The reason for these deviations remains unclear since all known systematic differences, such as the definition of cluster radii and the different energy bands used, were corrected for. These deviations therefore imply that there are additional systematic differences between the samples. However, for the central goal of this work, constraining the redshift evolution of scaling relations, this open question is of negligible importance because systematic differences mostly occur for low-redshift samples and the choice of sample from which multiply analyzed clusters are taken has no significant influence on the evolution results. In addition to systematic trends, even for samples that show no trends at all, the differences between the results considerably exceed the estimated errors. This indicates that the error estimations made by the different studies are too optimistic or that there are additional sources of measurement errors not included in the error budget.

Similar but less significant systematic trends are also visible when comparing the results for cluster masses. As visible in the top panel of Fig.\,\ref{siglowz}, $51\%$ of the results deviate by less than $1\sigma$ from the mean value, while $89\%$ lie within $2\sigma$ and no measurement shows deviations of more than $5\sigma$. However, apart from these systematic trends the estimated errors for cluster mass seem realistic as most measurements deviate by less than $1\sigma$ from the mean. The masses derived in Pratt09 based on the $Y_X$-parameter and the $Y_X$-M relation show no significant systematic difference from the hydrostatic mass estimates. However, owing to the small overlap sample of five clusters, the comparison analysis provides no suitable tool to identify these differences.  

In Fig.\,\ref{fig:sharedhighz}, the derived cluster properties of the $z\!>\!0.3$ overlap sample are compared. The typical observational errors for distant systems are larger, although within these errors the results are in closer agreement than for the local overlap sample.

The deviations from the mean temperature plotted in the bottom panel of Fig.\,\ref{sighighz} lie below the estimated errors for most systems. In detail, $72\%$ of the results deviate by less than $1\sigma$ and $90\%$ by less than $2\sigma$ from the mean value, while no deviations of more than $5\sigma$ occur. As for the low-redshift clusters, the measured temperatures show no systematic trends for single subsamples, \ie \ the spectroscopic fitting procedure also seems reliable for distant clusters and there appears to be no major systematic effects that have to be corrected. Furthermore, according to the mostly small deviations in units of the assumed error, the estimated error budget is likely to be realistic.   

Luminosities agree on average more strongly for the high-z clusters than for the local sample, for instance the Ettori04 and OHara07 results are consistent for 12 of the 18 clusters in common (see middle panel of Fig.\,\ref{fig:sharedhighz}). In contrast to the local overlap sample, no significant systematic trends between the different studies are visible. The deviations from the mean value plotted in the middle panel of Fig.\,\ref{sighighz} are smaller than $1\sigma$ for $53\%$ of the results and below $2\sigma$ for $62\%$ of the measurements. We have found that $12\%$ of the results deviate by more than $5\sigma$. The distribution of deviations implies that the error budget might have been previously underestimated by the different studies, although by no means as significantly as for the local systems.

The masses derived by Ettori04, Maughan06, and Kotov05 plotted in the top panel of Fig.\,\ref{fig:sharedhighz} are consistent within the errors for all shared clusters, all results deviate by less then $1\sigma$ from the mean value. The estimated errors are therefore likely realistic. However, the small size of the overlap sample of only four clusters does not allow us to peform a robust analysis of the systematic differences between the different studies.

\section{Local scaling relations for the combined cluster sample}
\label{sec:LocalScalingRelationsForTheCombinedClusterSample}

\begin{figure}[ht]
	\centering
		\includegraphics[angle=-90,width=0.5\textwidth]{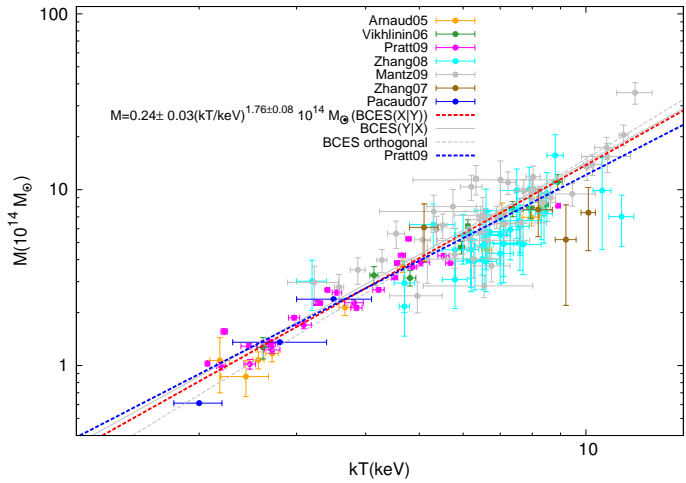}
	\caption{Local cluster sample: M--T relation. The red line shows the BCES(T$|$M) best-fit relation for the combined cluster sample, and the grey lines the BCES(M$|$T) and BCES orthogonal relations. The blue line shows the Pratt09 relation (see Sect.\,\ref{sec:LocalScalingRelations}).}
	\label{mtlowz}
\end{figure}

\begin{figure}[ht]
	\centering
		\includegraphics[angle=-90,width=0.5\textwidth]{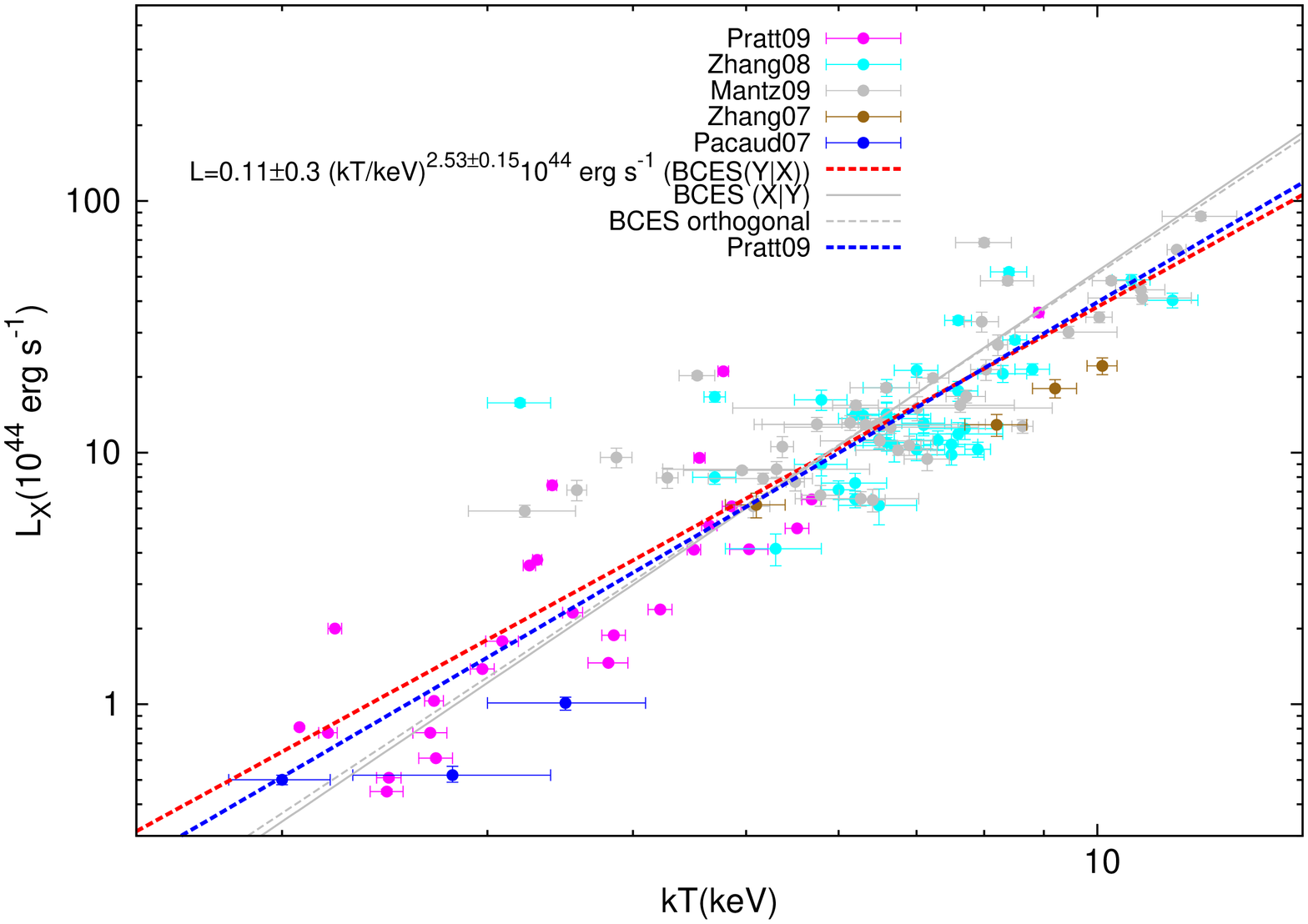}
	\caption{Local cluster sample: L$_X$--T  relation. The red line shows the BCES(L$|$T) best-fit relation for the combined cluster sample, and the grey lines the BCES(T$|$L) and BCES orthogonal relations. The blue line shows the Pratt09 relation (see Sect.\,\ref{sec:LocalScalingRelations}).}
	\label{ltlowz}
\end{figure}

\begin{figure}[ht]
	\centering
		\includegraphics[angle=-90,width=0.5\textwidth]{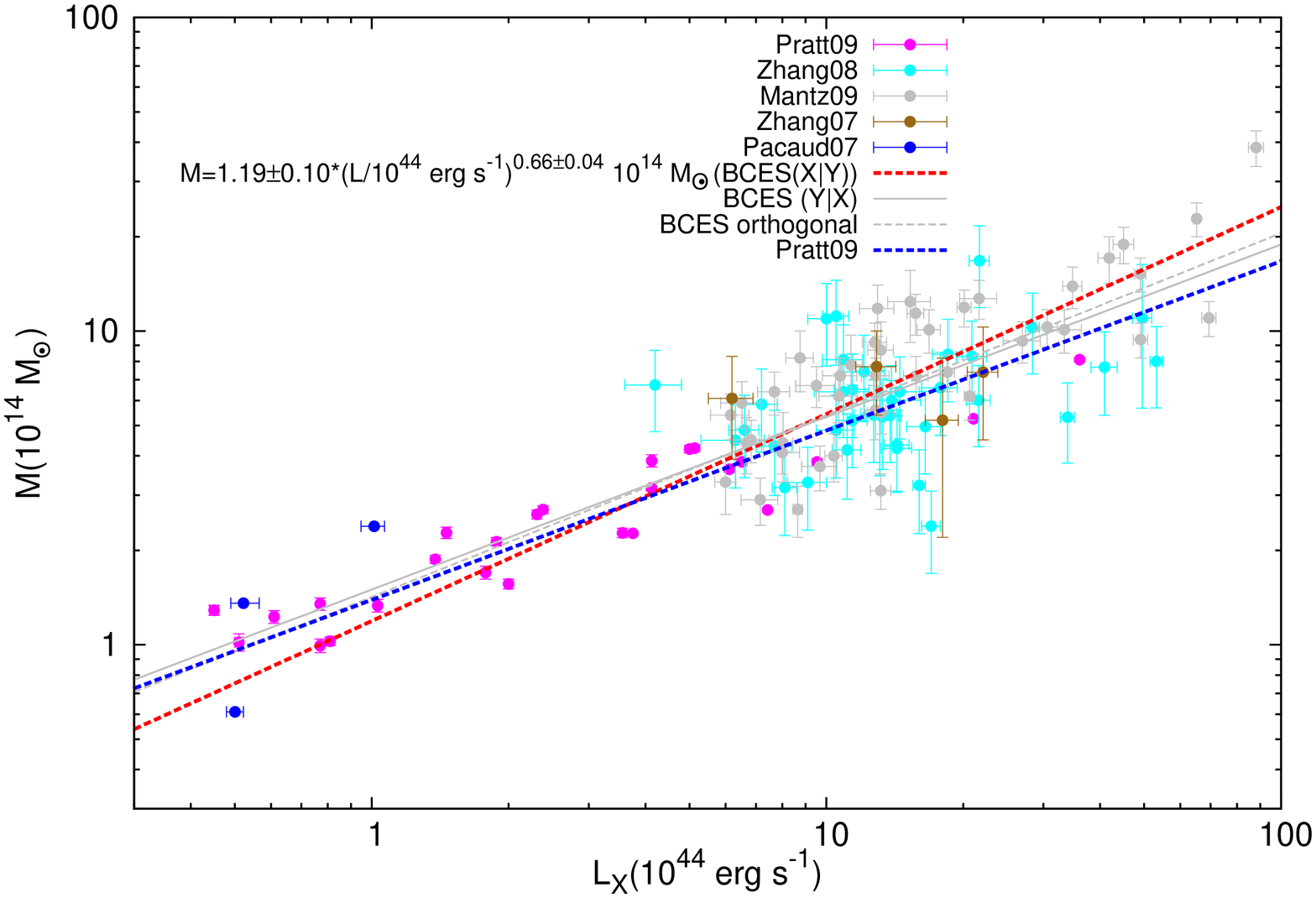}
	\caption{Local cluster sample: M--L$_X$ relation. The red line shows the BCES(L$|$M) best fit relation from the combined cluster sample, the grey lines the BCES(M$|$L) and BCES orthogonal relations. The blue line shows the Pratt09 relation (see Sect.\,\ref{sec:LocalScalingRelations}).}
	\label{mllowz}
\end{figure}

Fig.\,\ref{mtlowz}, \ref{ltlowz}, and \ref{mllowz} show the $z\!<\!0.3$-clusters included in the combined cluster samples and the local scaling relations fitted to this sample using different BCES fitting schemes in comparison to the relations derived by Pratt09 which were adopted for the evolution study in our work.

\end{appendix}
\end{document}